\documentclass{jfm}
\usepackage[a4paper]{geometry}
\usepackage{graphicx}
\usepackage{epstopdf,epsfig}
\usepackage{newtxtext}
\usepackage{newtxmath}
\usepackage{natbib}
\usepackage{hyperref}
\usepackage{booktabs}
\usepackage{array}
\usepackage{multirow}
\hypersetup{
    colorlinks = true,
    urlcolor   = blue,
    citecolor  = black,
}

\usepackage{setspace} 

\newcommand{\RomanNumeralCaps}[1]

\title{Onset of vortex shedding in flow past Rankine ovals}

\author{Zhaoyue Xu\aff{1,3},
  Yi Liu\aff{1,2}, Hua-dong Yao\aff{3},
 Shizhao Wang\corresp{\email{wangsz@lnm.imech.ac.cn}}\aff{1,2},
 \and Guowei He\aff{1,2}}

\affiliation{\aff{1} The State Key Laboratory of Nonlinear Mechanics, Institute of Mechanics, Chinese Academy of Sciences, Beijing 100190, China \\
\aff{2}School of Engineering Sciences, University of Chinese Academy of Sciences, Beijing 101408, China \\
\aff{3}Department of Mechanics and Maritime Sciences, Chalmers University of Technology, Gothenburg 41296, Sweden}
\begin{document}
\maketitle
\begin{abstract}
The Rankine oval is a classical geometry in potential flow, formed by superimposing a uniform stream with velocity $U$ and a source-sink pair separated by distance $2a$ with strength $m$, resulting in a closed stagnation streamline whose shape is governed by the dimensionless parameter $Ua/m$.
Although the Rankine body serves as a cornerstone for the classical theory of potential flow, its behavior in viscous flow remains unexplored.
The Rankine oval is streamlined in inviscid flow but behaves as a bluff body in viscous flow.
The onset of vortex shedding is a critical phenomenon in flows past a bluff body, mapping the transition from steady to periodic wakes.
This study systematically investigates the onset of vortex shedding in Rankine oval flows and its associated fluid dynamics by performing direct numerical simulations of incompressible flow past Rankine ovals over Reynolds numbers from 10 to 200 and $Ua/m$ from 0 to 1.
The investigation reveals a linear relationship between $Ua/m$ and the critical Reynolds number.
This study further characterizes the lift and drag coefficients and Strouhal number, analyzes the vortex formation, and performs a data-driven dimensional analysis.
This analysis identifies the dimensionless quantities and empirical formula that determine St and the friction drag coefficient as a function of Re, independent of $Ua/m$.
For sufficiently large $Ua/m$, the pressure drag can be estimated using potential flow solutions, enabling reliable predictions of the total drag without numerical simulations.
These conclusions collectively provide insights into the fluid dynamics of Rankine ovals across diverse flow conditions. 
\end{abstract}


\section{Introduction}
The Rankine oval, first introduced by William John Macquorn Rankine in the 19th century, stands as a classical solution in potential flow theory~\citep{Rankine1871, Wisniak2007}. 
This model is widely featured in fluid mechanics textbooks due to its elegance in demonstrating the superposition principle\citep{Batchelor1967, White1998}.
The Rankine oval demonstrates the application of the superposition principle by simply combining a uniform flow with a source-sink pair.
As an analytically solvable model, the Rankine oval allows for precise calculation of flow field parameters, such as velocity and pressure distributions, which is used in theoretical research to validate numerical methods or explore conservation laws in fluid mechanics.

Building on these theoretical foundations, the Rankine oval extends to a wide range of applications.
The Rankine oval serves as a fundamental model for analyzing the fluid dynamics of underwater hulls~\citep{Newman1977, Wang2020La, Liu2017}.
The accurate prediction of the flow field around a Rankine oval is essential, whether it moves at or beneath a free surface~\citep{SAHIN1993}.
The navigation of the Rankine oval in a submerged state starting at constant speed was analyzed by~\citet{bhlitem1966}. 
Furthermore,~\citet{Yu2022} numerically simulated the Rankine oval that represents underwater vehicle navigation, further exploring its acceleration and deceleration in stratified flow.
Indeed, the submerged bodies are chosen as Rankine ovals, representing a "building block" from which more general solutions can be derived~\citep{SAHIN1993}.
The hydrodynamic design of a tunnel vehicle is modeled on a Rankine oval incorporating a hole, as described by~\citet{SUNER20152}.
The significance of the Rankine oval shaped tunnel vehicle is analyzed in~\citet{Suner2018}.
In the work of~\citet{Suner2015}, the Rankine oval is employed to analyze the flows around bodies moving in both free and restricted fluid media.
This analysis is crucial for designing models that are energy-efficient and functionally effective.
In the field of wind energy,~\citet{Araya2014} introduced the Rankine oval as a representation for the vertical-axis wind turbines.
The Rankine oval is also applied in the field of acoustics. ~\citet{Olivier1998,Laik2000} used the Rankine ovals to study the acoustic scattering effects.
These models are useful for understanding acoustic wave behavior around aircraft bodies.
Moreover, numerous other studies have utilized Rankine ovals as models~\citep{GOEL1995,Xu2006,Zubarev1984}.

While Rankine ovals have found extensive application across various fields, the fundamental fluid dynamics of the flows around them remain largely unexplored.
In contrast, unsteady incompressible flows around other bodies, such as circles, ellipses, triangles, and rectangles, have attracted attention from researchers due to the complex physics underpinning these flows~\citep{LEKKALA2022}.
Among these, the flow around the circular cylinder—a canonical shape in fluid dynamics—has garnered  attention, serving as a benchmark.
Investigations of flow around the circular cylinder can be traced back to classical works by~\citet{Karman1911} and~\citet{Thom1933}, with numerous experimental and numerical analyses conducted since then~\citep{Roshko1961,Chorin1973,chew1995,chen1995,NORBERG2003,Lehmkuhl2013,wanga2018}.
Efforts have also been directed towards understanding the flow dynamics of other geometries, owing to their widespread applications in aerospace, civil, mechanical, and offshore engineering. 
For instance, \citet{MCLAREN1969} demonstrated through wind tunnel experiments that the drag coefficient of bluff sharp-edged cylinders, such as squares, is primarily dependent on turbulence intensity and Reynolds number.
\citet{xu2017} explored flow features of polygonal cylinders, while~\citet{LYSENKO2021,LYSENKO20211} delved into the flows around triangular and semi-circular cylinders, respectively.
These studies collectively revealed that drag coefficients, Strouhal numbers, and flow separation characteristics in such non-circular cylinders are highly dependent on polygon side number, geometric orientation and Reynolds number.
Additionally, the field of unsteady flows around bluff bodies has seen substantial contributions from other researchers, with early foundational works by~\citep{ZDRAVKOVICH1981,Bearman1984,Griffin1991,Williamson1996,Zdravkovich1997} focusing on vortex shedding, wake dynamics, and flow control, complemented by recent studies such as~\citep{thompson2014,DERAKHSHANDEH2019,LEKKALA2022} that explore wake transitions and high Re turbulence effects.
These investigations collectively offer a nuanced view of flow dynamics across various bluff bodies, highlighting the need for similar detailed studies on Rankine ovals to address the existing research gap.

To bridge this gap, it is essential to first review the primary characteristics of flows around canonical bluff bodies, such as cylinders, which have been extensively studied and provide a foundation for understanding more complex shapes like Rankine ovals. 
The key flow characteristics essential for understanding fluid dynamics around these bodies include flow separation, boundary layer development, vortex pattern dynamics, wake formations, and vortex shedding frequency~\citep{LEKKALA2022}.
As the Reynolds number increases, the flow—initially adhering to the surface of the body—separates, forming a shear layer.
At low Re, these vortices remain symmetric and stable in the wake.
Subsequently, as Re increases, the wake becomes unstable, resulting in the periodic shedding of alternating vortices in an anti-symmetric pattern, which is a phenomenon known as the Kármán Vortex Street~\citep{Karman1911}.
This transition from a steady to an unsteady wake occurs at a critical Reynolds number, $Re_c$, and can be characterized by a Hopf bifurcation~\citep{provansal1987,Noack1994}.
There has been longstanding interest in determining $Re_c$ for bluff bodies to better understand flow instabilities.
Although reported $Re_c$ values for a given geometry vary across studies, they typically cluster closely together.
As noted by~\citet{Williamson1996}, $Re_c$ is sensitive to factors such as subtle geometric variations, end conditions, and flow disturbances. 

Exceeding the critical Reynolds number $Re_c$ triggers vortex shedding, influencing key flow parameters like the Strouhal number and attracting research on various bluff bodies.
Foundational reviews~\citep{ZDRAVKOVICH1981,Bearman1984,Griffin1991} have established vortex shedding characteristics in bluff bodies, setting the stage for subsequent studies on gradually modified geometries and flow parameters~\cite{DERAKHSHANDEH2019}.
For instance,~\citet{thompson2014} numerically studied elliptic cylinders with aspect ratios ranging from 1 to 0, observing that $Re_c$ and $St_c$ increase as the geometry transitions from a flat plate to a circular cylinder.
\citet{Zafar2019} numerically investigated flow structures and heat transfer for cylinders modified from square to circular via corner radius changes at Re = 1000, observing increases in Strouhal number and Nusselt number with increasing radius ratio.
Similarly,~\citet{ABDELHAMID2021} examined the effects of corner radius on cylinders transitioning from square to circular, finding that $Re_c$ decreases from 49.5 to 46.7 with such modifications.
\citet{RASTAN2022} numerically investigated time-resolved laminar flow over rectangular cylinders with cross-sectional aspect ratios varying from square to flat plate at $Re = 40-100$, determining corresponding $Re_c$ values.
Furthermore, \citet{Chiarini2022} proposed a scaling law to predict the onset of primary Hopf instability in flows past two-dimensional bluff bodies, including ellipses, rectangles, triangles, and diamonds.

Rankine ovals are geometric bodies defined by variable parameters with clear physical interpretations in fluid dynamics, such as source-sink strength and separation distance.
While potential flow for Rankine ovals has been extensively explored, studies on their viscous flow characteristics and vortex shedding behavior under varying parameters remain limited.
To bridge this gap, the present numerical study investigates the onset and evolution of vortex shedding of Rankine ovals across a range of Reynolds numbers and geometric parameters, providing insights into how $Re_c$, $St$ and other characteristics vary with shape modifications.
Overall, this work aims to advance our understanding of Rankine oval wakes in unsteady flows.
The remainder of the paper is organized as follows:
Section 2 introduces and validates the numerical methods; Section 3 presents a detailed analysis of the results; and Section 4 draws the conclusions.

\section{Model and Numerical Methods}
\subsection{Rankine Oval}
\begin{figure}
  \centering
  \includegraphics[width=0.9\textwidth]{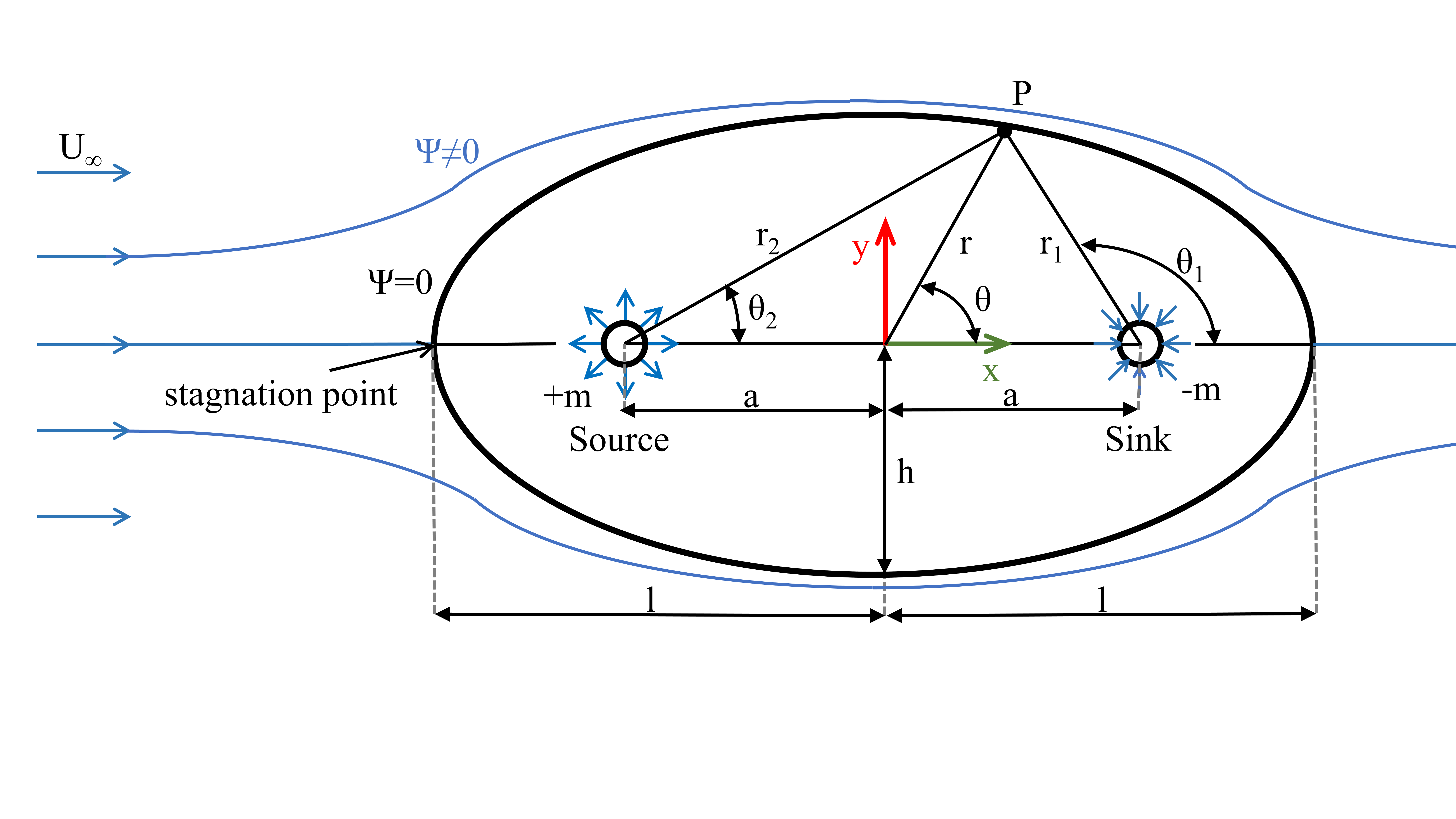}
  \caption{Schematic diagram of the geometric parameters of the Rankine oval. This figure illustrates the physical parameters involved in the manuscript. The flow field is the superposition of a source with strength $+m$ located at $(x, y)=(-a, 0)$, a sink with strength $-m$ located at $(x, y)=(a, 0)$, and a uniform stream with velocity $U_\infty$. The flow is described using the stream function $\psi$, with the Rankine oval located at the domain where $\psi = 0$. The Rankine oval intersects the y-axis at points $(0,h)$ and $(0,-h)$, and the x-axis at points $(l,0)$ and $(-l,0)$. The parameters $r_1$ and $\theta_1$ represent the distance and angle between an arbitrary point $P$ and the sink, while $r_2$ and $\theta_2$ represent the distance and angle between $P$ and the source. The distance and angle between $P$ and the origin are represented by $r$ and $\theta$, respectively.}
\label{fig1}
\end{figure}

The Rankine oval is a closed streamline that encloses a source and sink pair in a uniform free stream, as depicted in Fig.~\ref{fig1}. 
The coordinate system used in this work is established with its origin located at the centroid of the Rankine oval.
The x-axis is aligned parallel to the direction of the incoming uniform fluid flow, while the y-axis is oriented perpendicularly to the flow direction.
The source and sink are symmetrically positioned on the x-axis, each at a distance of $a$ from the origin, and possess equal but opposite strengths of $m$.
In this configuration, the stagnation points emerge at $x = \pm l$ as a result of the interaction between the upstream flow $U_\infty$ and the source-sink pair.
The Rankine oval is symmetric about both the x-axis and the y-axis.
The points of maximum width, where the tangent is parallel to the x-axis, occur at $(0, \pm h)$ on the y-axis.

The stream function for the superposition of potential flows is given by
\begin{equation}
\label{equ1}
    \psi =\frac{m}{2\pi}{\theta _2}-\frac{m}{{2\pi}}{\theta_1}+{U_\infty}y,
\end{equation}
where $\theta_1$ and $\theta_2$ are the angles from the x-axis to the line connecting point $P$ on the Rankine oval to the sink and source, respectively, as illustrated in Fig.~\ref{fig1}.
The first and second terms in Eq.~\eqref{equ1} correspond to the contributions from the source and sink, respectively, while the third term represents the uniform stream.

The shape of the Rankine oval is defined by the streamline $\psi=0$.
In Cartesian coordinates with axes x and y, the equation of the Rankine oval is given by
\begin{equation}
\label{RankineShap}
\frac{m}{2\pi}\tan^{-1}\left(\frac{y}{x+a}\right)-\frac{m}{2\pi}\tan^{-1}\left(\frac{y}{x-a}\right)+U_\infty y=0.
\end{equation}
The shape depends on the parameters $a$, $m$, and $U_\infty$.
The implicit form of Eq.~\eqref{RankineShap} is
\begin{equation}
\left(\frac{x}{a}\right)^2+\left(\frac{y}{a}\right)^2-1=\frac{2\frac{y}{a}}{\tan\left(2\pi\frac{U_\infty a}{m}\frac{y}{a}\right)}.
\end{equation}
As illustrated in Fig.~\ref{fig1}, the half body length $l$ is the distance from the origin to the stagnation points, while the half-height $h$ is half the extent along the y-axis.
The normalized half-length and half-height are given by
\begin{equation}
\frac{l}{a}=\left(\frac{1}{\pi}\frac{m}{U_\infty a}+1\right)^{\frac{1}{2}},
\frac{h}{a}=\cot{\frac{\frac{h}{a}}{2\frac{m}{U_\infty a}}}.
\label{halfLengthhalfWidth}
\end{equation}
The shape is determined by the non-dimensional parameter $U_\infty a/m$, which represents the inverse of non-dimensional source strength.
For the sake of clarity and brevity in our discussion, we  denote the non-dimensional number $U_\infty a/m$ as $Ua/m$ throughout this manuscript.
The term aspect ratio (AR) refers to the ratio of the length to the height, describing the elongation of the Rankine oval.
\begin{equation}
    AR=l/h
\end{equation}
Different values of $Ua/m$ produce Rankine ovals with varying AR.
Larger values of $Ua/m$ yield elongated, slender shapes, while smaller values approach a circular form.
To enable a standardized comparison of different Rankine ovals, we applied a normalization with $h=0.5$, ensuring that the vertical diameter $D$ is uniformly equal to $1$ for all shapes.
Table~\ref{tab1} lists $Ua/m$ values and corresponding geometric features.

\begin{table}
    \centering
    \caption{\label{tab1}Rankine oval geometry features at different $Ua/m$ values.}
    \begin{tabular}{c c c c c c}
    \hline
       $Ua/m$  & $l$ & $a$ & $h$ & $l/h$(AR) & $l/a$\\
    \hline
       0.0  & 0.500 & 0.000 & 0.500 & 1.000 & $\infty$\\
       0.1  & 0.603 & 0.295 & 0.500 & 1.206 & 2.045\\
       0.2  & 0.704 & 0.437 & 0.500 & 1.408 & 1.610\\
       0.3  & 0.804 & 0.560 & 0.500 & 1.607 & 1.436\\
       0.4  & 0.903 & 0.674 & 0.500 & 1.806 & 1.340\\
       0.5  & 1.002 & 0.783 & 0.500 & 2.004 & 1.279\\
       0.6  & 1.101 & 0.890 & 0.500 & 2.202 & 1.237\\
       0.7  & 1.200 & 0.995 & 0.500 & 2.400 & 1.206\\
       0.8  & 1.299 & 1.099 & 0.500 & 2.598 & 1.182\\
       0.9  & 1.398 & 1.202 & 0.500 & 2.796 & 1.163\\
       1.0  & 1.497 & 1.304 & 0.500 & 2.994 & 1.148\\
       \hline
    \end{tabular}
\end{table}

In Table~\ref{tab1}, we present the values of the half-length $l$, half-height $h$, and the half-distance $a$ between the source and sink pair, as well as the ratios $l/h$ and $l/a$ for various values of $Ua/m$.
Notably, the relationship between $Ua/m$ and $l/h$ can be approximated linearly, as detailed in Appendix A.
For $h=0.5$, the following approximation of the aspect ratio holds:
\begin{equation}
\label{Eqlh}
\frac{l}{h} \approx 1 + 2\frac{{Ua}}{m}.
\end{equation}
Table~\ref{tab1} and Eq.~\eqref{Eqlh} indicate that increasing the freestream flow speed or decreasing the source-sink strength results in a larger aspect ratio of the Rankine oval.

\subsection{Numerical setting}
The numerical simulation in this work is based on the incompressible Navier-Stokes (NS) equations,
\begin{equation}
\label{NS}
\begin{aligned}
    \frac{\partial{{u}_i}}{\partial {x_i}} & = 0, \\
    \frac{\partial {u}_i}{\partial {t}} + {u}_j\frac{\partial {u}_i}{\partial {x_j}} &= - \frac{\partial p}{\partial x_i} + \nu \frac{{\partial ^2}{u}_i}{\partial x_j\partial x_j},
\end{aligned}
\end{equation}
where $u_i$ denotes velocity components in the Cartesian coordinates $x_i\left( {i = 1,2,3} \right)$, representing the streamwise, vertical, and lateral directions, respectively.
For the 2D flows investigated in this work, we consider $i=1,2$, where $x_1$ and $x_2$ correspond to $x$ and $y$ in the coordinate system shown in Figure~\ref{fig1}.
Here, $p$ and $\nu$ represent the pressure and kinematic viscosity of the fluid, respectively.


The NS equations are discretized using the cell-centered finite volume method (FVM) on hybrid unstructured meshes.
We employ an in-house computational fluid dynamics solver to solve these equations~\citep{LIU201819}.
The discretized convective flux term is computed with the second-order Roe scheme, and the viscous flux term is obtained by the reconstructed central scheme.
For the Roe scheme, second-order accuracy is achieved by reconstructing the solution following Barth's interpolation method~\citep{barth1989design}.
A second-order fully implicit dual-time scheme is employed for the unsteady simulations.
To ensure computational stability and efficiency while mitigating the negative impact of poorly resolved grids, we adopt an adaptive local time-stepping method.
For large-scale computations, we parallelize the solver using a domain decomposition strategy facilitated through the Message Passing Interface (MPI) protocol.
Additionally, we utilize nonblocking communications to efficiently overlap computation with communication.
The detailed numerical methods and validation can be found in previous work~\citep{Liu2021,LIU2019}.

\begin{figure}
  \centering
  \includegraphics[width=1.0\textwidth]{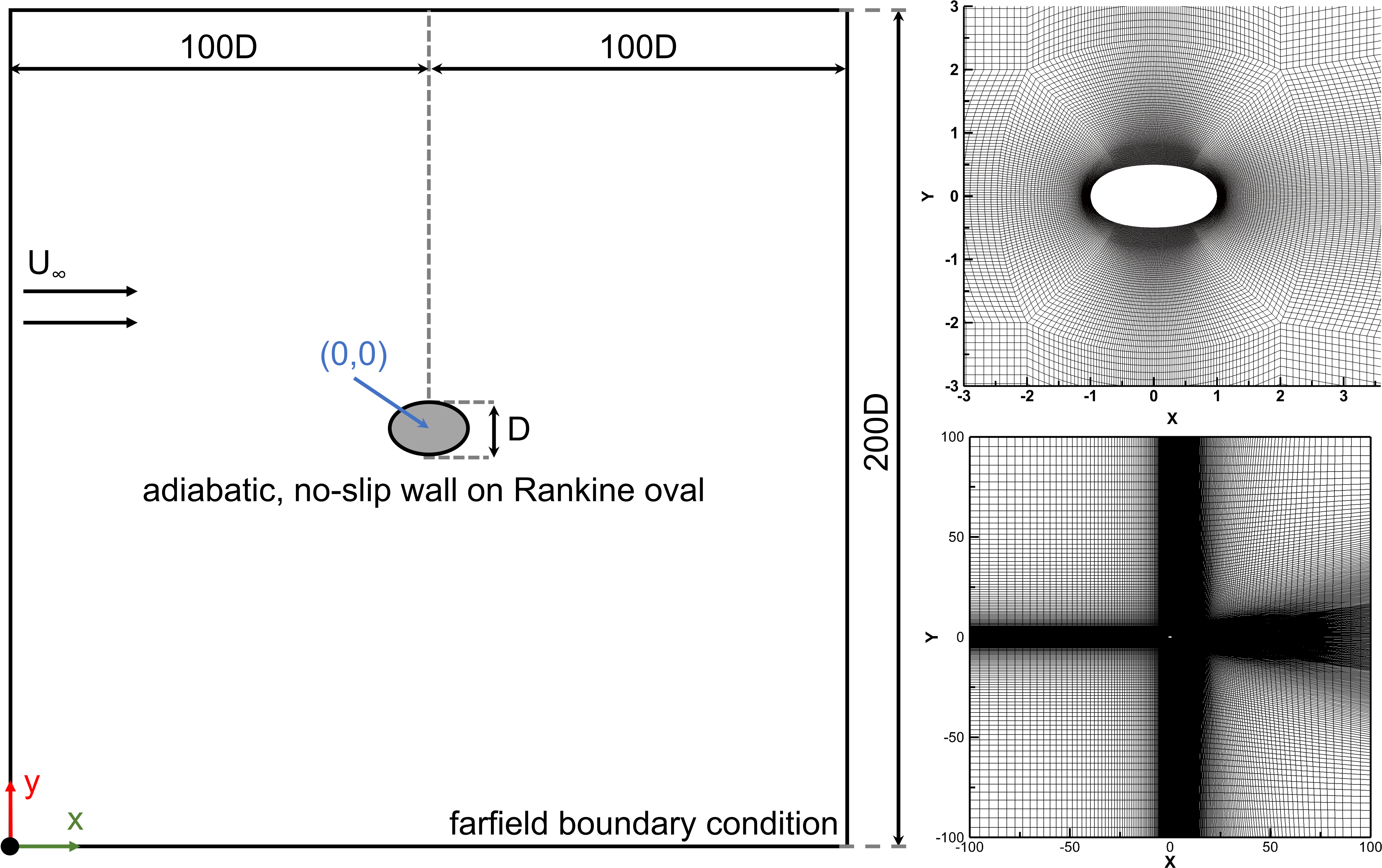}
  \caption{Schematic of the computational domain and mesh details for the Rankine oval simulation. The domain is a two-dimensional rectangular domain with clearly defined boundaries, and the figure highlights the mesh resolution and structure around the Rankine oval.}
  \label{Fig2}
\end{figure}
We simulate Rankine ovals in free-stream flow at zero angle of attack.
As shown in Figure~\ref{Fig2}, the simulations employ a square computational domain of $[-100D,100D]\times[-100D,100D]$, where $D = 1$ is the height of the Rankine oval.
This setup ensures a maximum blockage ratio of $0.005$.
The domain is sufficiently large to prevent far-field boundaries from affecting the results.
A free-slip boundary condition is applied at the far field, while a no-slip condition is used on the Rankine oval surface.
The first grid layer near all walls is less than $0.01D$, with a growth rate of $1.05$.
The total number of cells ranges from $78,000$ to $120,000$ depending on the value of $Ua/m$ from $0.0$ to $1.0$.
The nondimensional time step is $\Delta t^* = \Delta t U_\infty / D$.
The simulations begin with the flow field at rest, where $u = v = 0$ at $t^* = 0$.
Grid independence is discussed later in this section.

\begin{table}
\centering
\caption{\label{simulation_cases}Setups of simulation cases at different $Re$}
\begin{tabular}{c c p{11cm}}
\hline
$Ua/m$ & Case Nos. & Simulation Cases ($Re$) \\ \hline
0.0 & 1-18 & 10, 20, 30, 40, 42, 44, 46, 48, 50, 52, 55, 60, 70, 80, 90, 100, 150, 200 \\ 
0.1 & 19-38 & 10, 20, 30, 40, 50, 52, 54, 56, 58, 60, 62, 64, 66, 68, 70, 80, 90, 100, 150, 200 \\ 
0.2 & 39-58 & 10, 20, 30, 40, 50, 52, 54, 56, 58, 60, 62, 66, 68, 70, 72, 80, 90, 100, 150, 200 \\ 
0.3 & 59-78 & 10, 20, 30, 40, 50, 60, 62, 64, 66, 68, 70, 72, 74, 76, 78, 80, 90, 100, 150, 200 \\ 
0.4 & 79-101 & 10, 20, 30, 40, 50, 60, 64, 66, 68, 70, 72, 74, 76, 78, 80, 82, 84, 86, 88, 90, 100, 150, 200 \\ 
0.5 & 102-122 & 10, 20, 30, 40, 50, 60, 70, 72, 74, 76, 78, 80, 82, 84, 86, 88, 90, 92, 100, 150, 200 \\ 
0.6 & 123-141 & 10, 20, 30, 40, 50, 60, 70, 80, 82, 84, 86, 90, 92, 94, 96, 98, 100, 150, 200 \\ 
0.7 & 142-165 & 10, 20, 30, 40, 50, 60, 70, 80, 86, 88, 90, 92, 94, 96, 98, 100, 104, 106, 108, 110, 112, 120, 150, 200 \\ 
0.8 & 166-189 & 10, 20, 30, 40, 50, 60, 70, 80, 90, 92, 94, 96, 98, 100, 102, 104, 106, 108, 110, 112, 114, 120, 150, 200 \\ 
0.9 & 190-213 & 10, 20, 30, 40, 50, 60, 70, 80, 90, 96, 98, 100, 102, 104, 110, 118, 120, 122, 124, 126, 130, 140, 150, 200 \\ 
1.0 & 214-235 & 10, 20, 30, 40, 50, 60, 70, 80, 90, 100, 102, 104, 106, 108, 110, 120, 122, 124, 130, 140, 150, 200 \\ \hline
\end{tabular}
\end{table}
\begin{table}
\centering
\caption{\label{tab:grid_sensitivity}Grid and computational domain independence verification for the flow around a circular cylinder}
\begin{tabular}{c c c c c c c c c}
\hline
Name & Number of Grid Cells & Computational Domain & $\overline{Cd}$& diff \% & $\overline{Cd_p}$ & diff \% & $\overline{Cd_f}$ & diff \% \\
\hline
Coarse Grid & 19,118   & $200D\times200D$  & 1.50 & 1.3\% & 0.97 & 3.0\% & 0.53 & 1.9\% \\
Fine Grid & 325,632   & $200D\times200D$   & 1.51 & 0.6\% & 0.98 & 2.0\% & 0.53& 1.9\%  \\
Small Domain Grid   & 30,072 &$10D\times10D$ & 1.06 & 30.3\% & 0.75 & 25.0\% & 0.31& 40.4\% \\
Large Domain Grid & 328,176  & $400D\times400D$ & 1.51  & 0.6\% & 0.98 & 2.0\% & 0.53& 1.9\% \\
Selected Grid  & 77,172 & $200D\times200D$  & 1.51& 0.6\%  &0.98 & 2.0\%  & 0.53 &1.9\% \\
~\citet{Dennis1970} & – & –     & 1.52 & –     & 1.00 & –     & 0.52  \\
\hline
\end{tabular}
\end{table}
All the cases simulated are listed in Table~\ref{simulation_cases} for various values of $Ua/m$ and $Re$.
The key nondimensional parameters calculated in these simulations include the Strouhal number $St$, drag coefficient $Cd$, lift coefficient $Cl$, and pressure coefficient $Cp$, defined as follows\cite{RASTAN2022}:
\begin{align}
    Re &= \frac{{U_\infty}D}{\nu} \\
    St &= \frac{fD}{U_\infty} \\
    Cl &= \frac{F_l}{0.5\rho U_\infty^2D} \\
    Cd &= \frac{F_d}{0.5\rho U_\infty^2D} \\
    Cp &= \frac{p - {p_\infty}}{0.5\rho U_\infty^2}
\end{align}
where $F_d$, $F_l$, and $f$ represent the drag force, lift force and vortex shedding frequency, respectively.
The drag coefficient is further decomposed into the pressure drag coefficient $Cd_p$ and the friction drag coefficient $Cd_f$, and the overline denotes the time-averaged value.

Before presenting the results, tests are conducted to evaluate the sensitivity of hydrodynamic forces to grid resolution and computational domain size, ensuring independence in the computational fluid dynamics simulations.
Theoretically, when $Ua/m = 0$, the corresponding Rankine oval reduces to a circular cylinder, serving as an ideal benchmark for validation.
As illustrated in Table~\ref{tab:grid_sensitivity}, preliminary simulations are performed to assess the impact of grid resolution and domain size on the hydrodynamic forces.
The coarse and fine grids show minor differences in $\overline{Cd}$, $\overline{Cd_p}$, and $\overline{Cd_f}$. However, the results of the coarse grid are slightly less accurate than those of the selected and fine grids, leading to its exclusion.
The small domain grid, confined to a $10D \times 10D$ domain, exhibits significant deviations in drag coefficients, while the large domain grid, extending to $400D \times 400D$, offers negligible improvements over its $200D \times 200D$ counterpart. 
Consequently, the selected grid, which balances resolution and computational efficiency, is chosen.

To validate the simulations, the drag and lift coefficients are compared with those from previous studies, as shown in Table~\ref{tab:coefficients}. 
The results from the present study align closely with the reported values.
\begin{table}
\centering
\caption{\label{tab:coefficients}Comparison of drag and lift coefficients for flow past a cylinder at $Re = 100$ and $Re = 200$.}
\begin{tabular}{l c c c }
\hline
Case & Re & $\overline{Cd}$ & $\overline{Cl}$  \\
\hline
Present study & 100 & 1.33 & $\pm$0.33 \\
\cite{WANG20113} & 100 & 1.33 & $\pm$0.32  \\
\cite{LIU1998} & 100 & 1.35 & $\pm$0.32 \\
\cite{Park1998} & 100 & 1.33 & $\pm$0.33  \\
\cite{UHLMANN2005} & 100 & 1.45 & $\pm$0.34  \\
\hline
Present study & 200 & 1.34 & $\pm$0.69 \\
\cite{WANG20113} & 200 & 1.32 & $\pm$0.69  \\
\cite{LINNICK2005} & 200 & 1.34 & $\pm$0.69  \\
\cite{LIU1998} & 200 & 1.31 & $\pm$0.69  \\
\cite{TAIRA2007} & 200 & 1.35 & $\pm$0.68 \\
\hline
\end{tabular}
\end{table}

\subsection{Hopf bifurcation}
Hopf bifurcation occurs when a small change in the parameters of dynamic system leads to a shift in stability, typically from a stable fixed point to a limit cycle.
This bifurcation is evident in flows around bluff bodies, such as a cylinder, where it marks the onset of vortex shedding, signifying the transition from a steady to an unsteady state. 
The Stuart-Landau equation serves as a fundamental model in nonlinear dynamics for describing the behavior of a nonlinear oscillator near such a bifurcation.
This equation has been validated in various studies~\citep{THOMPSON2004,Rastan2021} and is given by
\begin{equation}
\label{Landau}
\frac{dA}{dt} = \sigma A - L|A|^2 A,
\end{equation}
where $A(t)$ is a complex variable representing the system state, $L$ is the Landau coefficient, and $\sigma = \sigma_r + i\sigma_i$ denotes the complex linear growth rate, with $\sigma_r$ as the amplification rate and $\sigma_i$ as the angular frequency of the oscillations.
As noted by~\citet{RASTAN2022}, the global coefficients $\sigma$ and $L$ remain constant throughout a specific flow.

\begin{figure}
  \centering
  \includegraphics[width=1.0\textwidth]{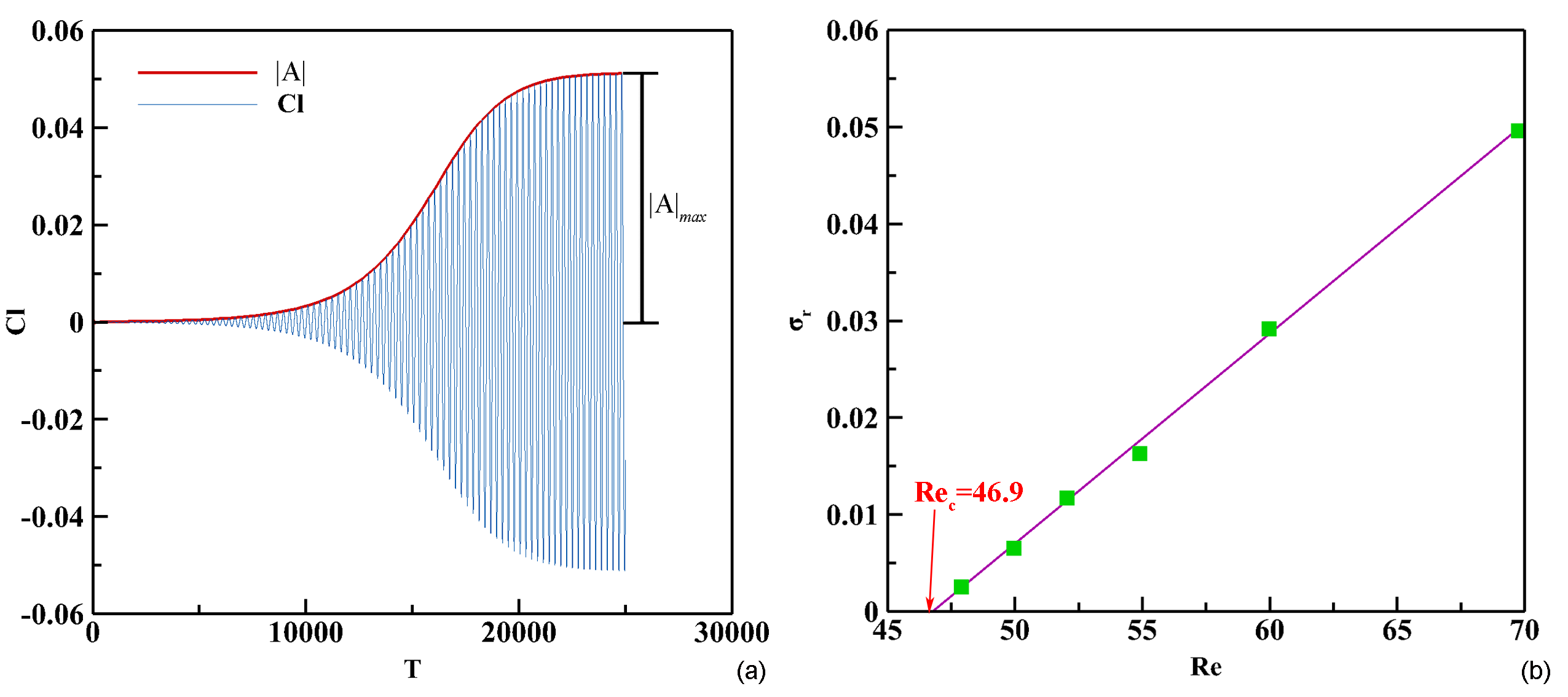}
    \caption{\label{LandauFianl}Schematic of the Stuart-Landau equation analysis: (a) Time history of the lift coefficient, and (b) linear global growth rate as a function of $Re$ to determine $Re_c$.}
\end{figure}

We specifically focus on the real growth rate $\sigma_r$ in Eq.~\eqref{Landau} to determine the critical Reynolds number $Re_c$.
This equation can be reformulated in terms of the real variables as follows\citep{RASTAN2022},
\begin{equation}
\frac{1}{|A|}\frac{d|A|}{dt} = \sigma_r\left(1 - \frac{|A|^2}{|A|_{max}^2}\right),
\end{equation}
where $|A|$ denotes the envelope amplitude of the state.
For the onset of vortex shedding, the amplitude $A(t)$ corresponds to the lift coefficient $Cl(t)$, and the real growth rate $\sigma_r$ is related to the Reynolds number $Re$.
For $Re < Re_c$, the negative growth rate suppresses the growth of $Cl(t)$, maintaining a non-oscillatory state.
Conversely, for $Re > Re_c$ where $\sigma_r > 0$, $Cl(t)$ exhibits oscillatory behavior.
Thus, the condition $Re = Re_c$ corresponds to $\sigma_r = 0$.
As shown in Fig.~\ref{LandauFianl}, by calculating $\sigma_r$ for several Reynolds numbers near $Re_c$ and fitting a line to the $\sigma_r$-$Re$ relationship, the critical Reynolds number $Re_c$ can be determined.

Our results can be validated by comparing the computed $Re_c$ for the circular cylinder with values extensively documented in the literature, as detailed in Table~\ref{tabReccylinder}.
\begin{table}
\setlength\tabcolsep{10pt}
    \centering
    \caption{\label{tabReccylinder}Summary of $Re_c$ and $St_c$ values at the onset of periodic vortex shedding for a circular cylinder.}
    \begin{tabular}{c c c c}
    \hline
       Case  & $Re_c$ & $St_c$ & Method \\
    \hline
        Present study & 46.9 & 0.115 & Numerical Simulation \\
        \citet{jackson1987}  & 46.2 & 0.138 & Numerical Simulation \\
       \citet{williamson1989}  & 47.9 & 0.122 & Experiment Measurement\\
       \citet{MORZYNSKI1999}  & 47.0 & 0.132 & Numerical Simulation \\
       \citet{KUMAR2006}  & 46.9 & 0.117 & Numerical Simulation \\
       \citet{thompson2014}  & 47.2 & 0.116 & Experimental Measurement \\
       \citet{leontini2015}  & 46.6 & 0.119 & Numerical Simulation \\
        \citet{Chopra2019}  & 47.0 & 0.114 & Numerical Simulation \\
        \citet{ABDELHAMID2021}  & 46.7 & 0.113 & Numerical Simulation \\
       \hline
    \end{tabular}
\end{table}

Analysis of the data in Table~\ref{tabReccylinder} shows that the average $Re_c$ is 46.9, ranging from 46.2 to 47.9, while $St_c$ averages 0.12 and ranges from 0.11 to 0.14.
In our simulations, as shown in Fig.~\ref{LandauFianl}, $Re_c$ is 46.9 and $St_c$ is 0.115, which align closely with the literature values, confirming the validity of our simulation method for determining the critical Reynolds number.

\section {Results and Discussion}
This section presents the numerical results from simulations of flow around Rankine ovals, with a detailed analysis of key findings on flow patterns, vortex behavior, and variations in dimensionless quantities.
These results highlight the complex dynamics of vortex shedding and their implications in fluid dynamics.

Section\ref{subsec:criticalReynoldsNumber} investigates the $Re_c$ for flow around Rankine ovals, establishing its linear relationship with the parameter $Ua/m$ and comparing it across different geometries.
Section\ref{subsec:flowFieldAndDimensionlessParametersAtCriticalReynoldsNumber} examines the flow fields, vortex structures, and dimensionless quantities near the critical Reynolds number.
Section\ref{subsec:Pressure Distributions from Potential Flow and Numerical Simulation} compares pressure distributions from numerical simulations with potential flow theory, assessing approximations for pressure drag at high $Ua/m$.
Section\ref{subsec:dragCoefficientsAndStrouhalNumbersAcrossReynoldsNumbersAndRankineGeometries} analyzes variations in drag coefficients (including pressure and friction components) and Strouhal numbers across a range of Reynolds numbers and Rankine oval shapes.
Section\ref{subsec:vortexSheddingStructure} explores the structures of vortex shedding under varying $Re$ and $Ua/m$, grouping similar wake patterns.
Section\ref{subsec:velocityFluctuations} evaluates velocity fluctuations, focusing on root-mean-square distributions, vortex formation lengths, and wake widths.
Section\ref{subsec:instantaneousLiftAndDragCoefficients} details the instantaneous lift and drag coefficients, their amplitudes, and decompositions into pressure and friction components.
Finally, Section\ref{subsec:dimensionanalysis} applies data-driven dimensional analysis to identify key dimensionless quantities governing friction drag and Strouhal number, independent of $Ua/m$, offering a new perspective for understanding and predicting the flows around Rankine oval.

\subsection{\label{subsec:criticalReynoldsNumber}Critical Reynolds Number}
This subsection focuses on the critical Reynolds number ($Re_c$) for flow around Rankine ovals, investigating its relationship with the geometric parameter $Ua/m$ to reveal the underlying dynamics.

\begin{figure}
  \centering
  \includegraphics[width=1.0\textwidth]{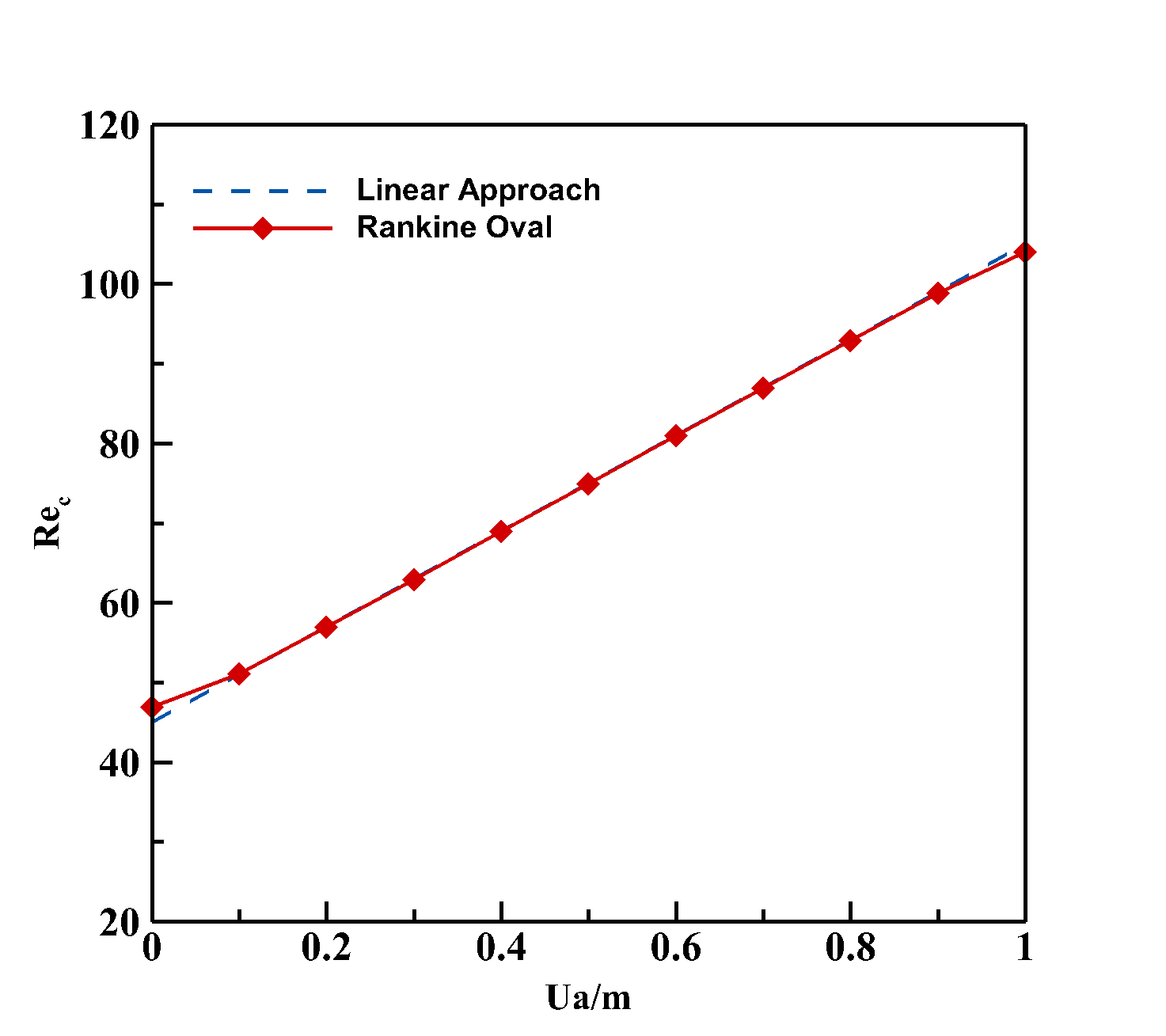}
  \caption{The relationship between the critical Reynolds number $Re_c$ and the parameter $Ua/m$ for Rankine ovals. Each case is represented by a red diamond, connected by a red line. The dashed blue line represents the linear equation $Re_c = 60Ua/m + 45$.}
\label{fig5}
\end{figure}

Figure~\ref{fig5} illustrates the relationship between $Re_c$ and $Ua/m$, revealing an approximately linear trend derived from extensive numerical simulations.
As $Ua/m$ increases from 0 to 1, $Re_c$ rises correspondingly, from approximately 46.9 (for a circular geometry at $Ua/m = 0$) to 104.0 (at $Ua/m = 1.0$).
This trend is modeled by the linear equation $Re_c = 60Ua/m + 45$, which provides a reliable approximation within the investigated range of $0.1 \leq Ua/m \leq 1.0$.
Minor deviations occur at the boundaries at $Ua/m = 0$.
Despite these discrepancies, the linear model effectively captures the behavior for the geometric parameters studied.

\begin{figure}
  \centering
  \includegraphics[width=1.0\textwidth]{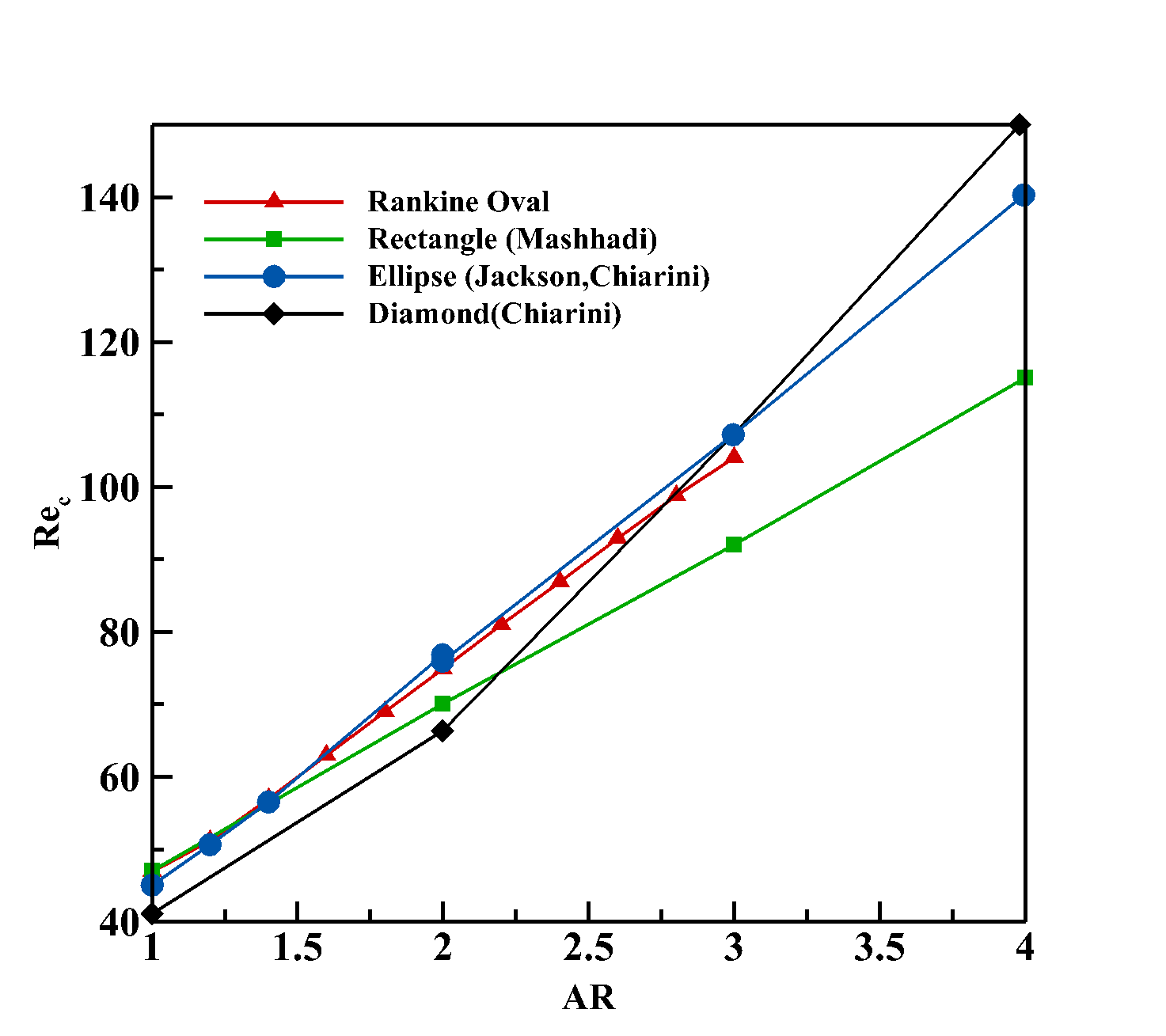}
  \caption{Comparison of the critical Reynolds number $Re_c$ for Rankine ovals, rectangles, diamonds, and ellipses across varying AR. Rankine oval cases are denoted by red triangles connected with a red line. Rectangular cases, from\citet{MASHHADI2021}, are indicated by green rectangles connected with a green line. Elliptical cases, based on\citet{jackson1987} and\citet{Chiarini2022}, are represented by blue circles connected with a blue line. Diamond cases, from\citet{Chiarini2022}, are depicted by black diamonds connected with a black line.}
\label{figRecCompare}
\end{figure}
To contextualize these findings, Fig.~\ref{figRecCompare} compares $Re_c$ across different geometries (Rankine ovals, rectangles, diamonds, and ellipses) as a function of AR. In the range $1.0 \leq AR \leq 1.5$, $Re_c$ values are similar for Rankine ovals, rectangles, and ellipses.
However, as AR increases beyond this, significant divergence emerges.
Rankine ovals and rectangles~\citep{MASHHADI2021} exhibit linear increases in $Re_c$, while ellipses and diamonds display nonlinear growth with large increments.

\subsection{\label{subsec:flowFieldAndDimensionlessParametersAtCriticalReynoldsNumber}Flow Fields and parameters for critical Reynolds number}
\begin{figure}
  \centering
  \includegraphics[width=0.97\textwidth]{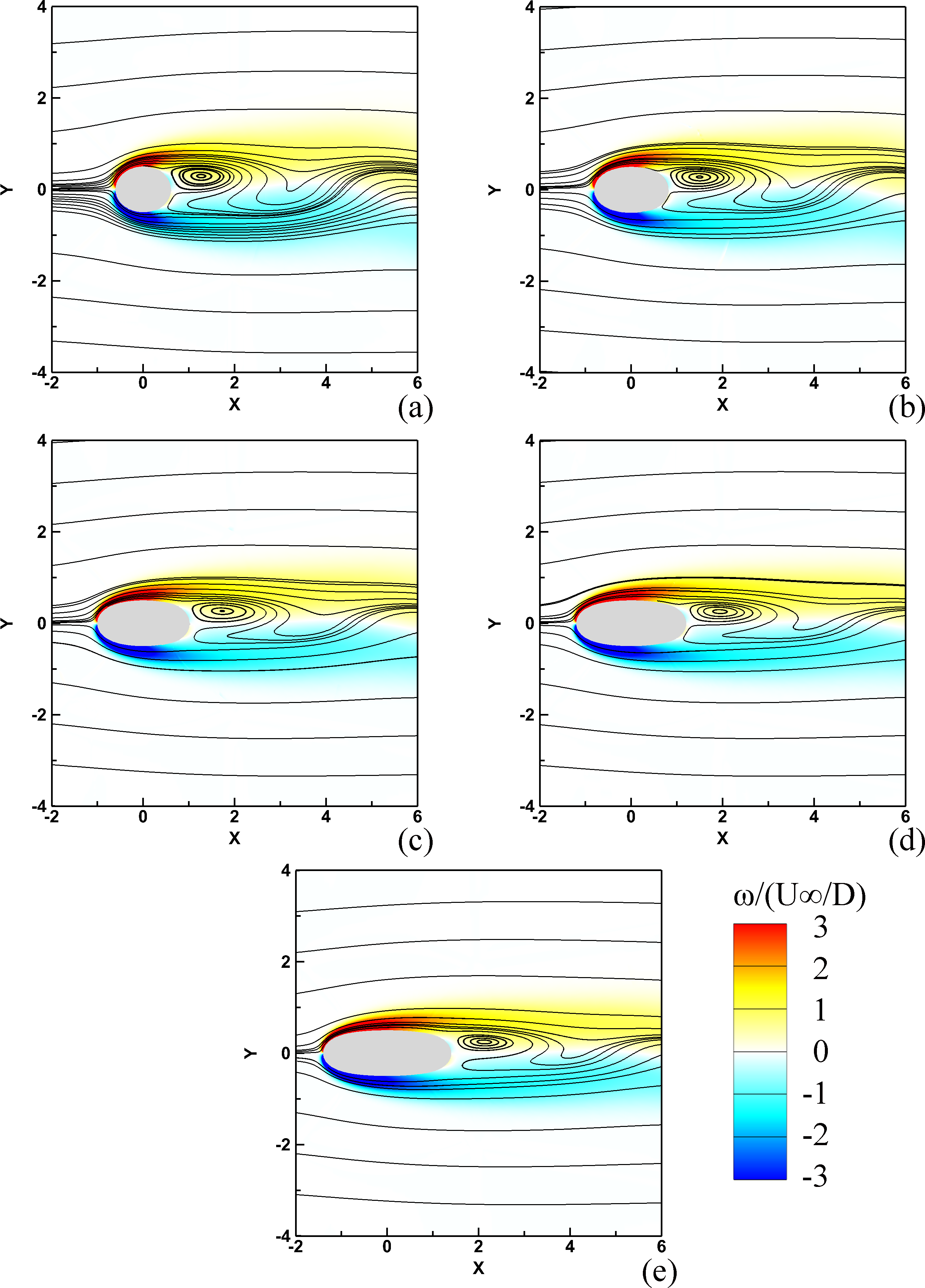}
  \caption{Instantaneous flow structure, streamlines, and vorticity for Rankine ovals with (a) $Ua/m=0.1$, (b) $Ua/m=0.3$, (c) $Ua/m=0.5$, (d) $Ua/m=0.7$, and (e) $Ua/m=0.9$. Each subfigure corresponds to the state of maximum lift coefficient as the flow approaches the critical Reynolds number $(Re - Re_c < 2)$.
  }
 \label{fig3}
\end{figure}

\begin{figure}
  \centering
  \includegraphics[width=0.97\textwidth]{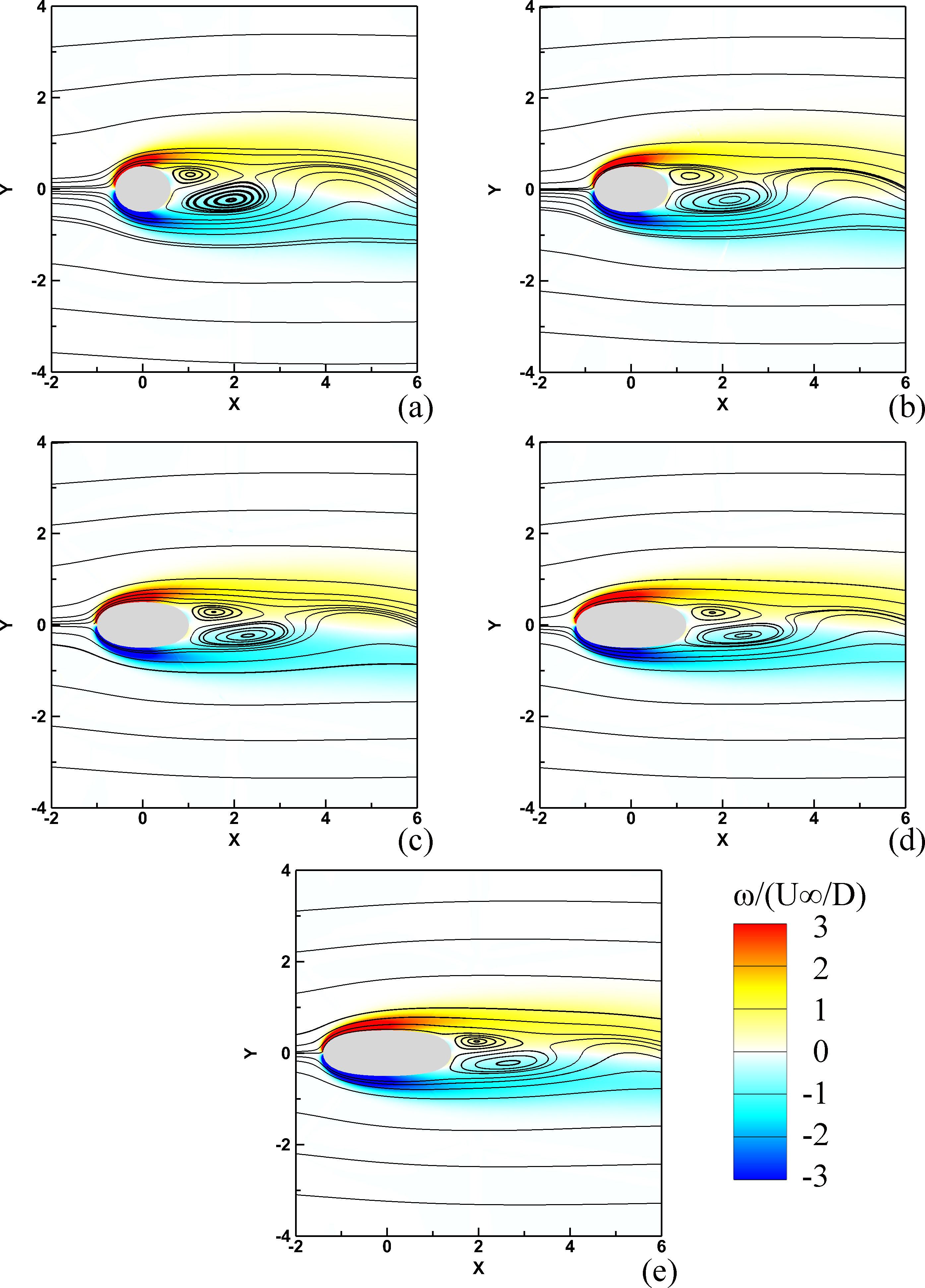}
  \caption{Instantaneous flow structure, streamlines, and vorticity for Rankine ovals with (a) $Ua/m=0.1$, (b) $Ua/m=0.3$, (c) $Ua/m=0.5$, (d) $Ua/m=0.7$, and (e) $Ua/m=0.9$. Each subfigure corresponds to the state of zero lift coefficient as the flow approaches the critical Reynolds number $(Re - Re_c < 2)$.
  }
\label{fig4}
\end{figure}

Following the establishment of the relationship between the $Re_c$ and the parameter $Ua/m$ for Rankine ovals, this subsection presents visualizations of the flow fields around Rankine ovals as they approach the critical Reynolds number.
The flow fields at specific phases of the lift coefficient are illustrated in Figs.~\ref{fig3} and~\ref{fig4}.

These figures offer insights into the flow behavior around Rankine ovals near the critical Reynolds number. 
Streamlines are shown as black solid lines, the Rankine oval in gray, and vorticity contours from blue to red with $\omega/(U_{\infty}D) \in [-3, 3]$, where positive values dominate above the oval and negative values below.
The flow structure exhibits a similar pattern in the trailing edge vortex.
At the phase corresponding to the maximum lift coefficient, the flow features a single distinct vortex bubble.
In contrast, two vortex bubbles are observed when the lift coefficient is zero.
The upper and lower vortex bubbles form successively in the wake of the Rankine oval and convect downstream.
After the dissipation of a developed large vortex bubble, a temporal delay occurs before a new, smaller vortex bubble forms, leading to a phase with only one vortex bubble present.
Additionally, the flow field exhibits asymmetry as the lift coefficient approaches zero, with the upper vortex bubble being smaller than the lower one, highlighting the gradual development of this asymmetry.

\begin{figure}
  \centering
  \includegraphics[width=1.0\textwidth]{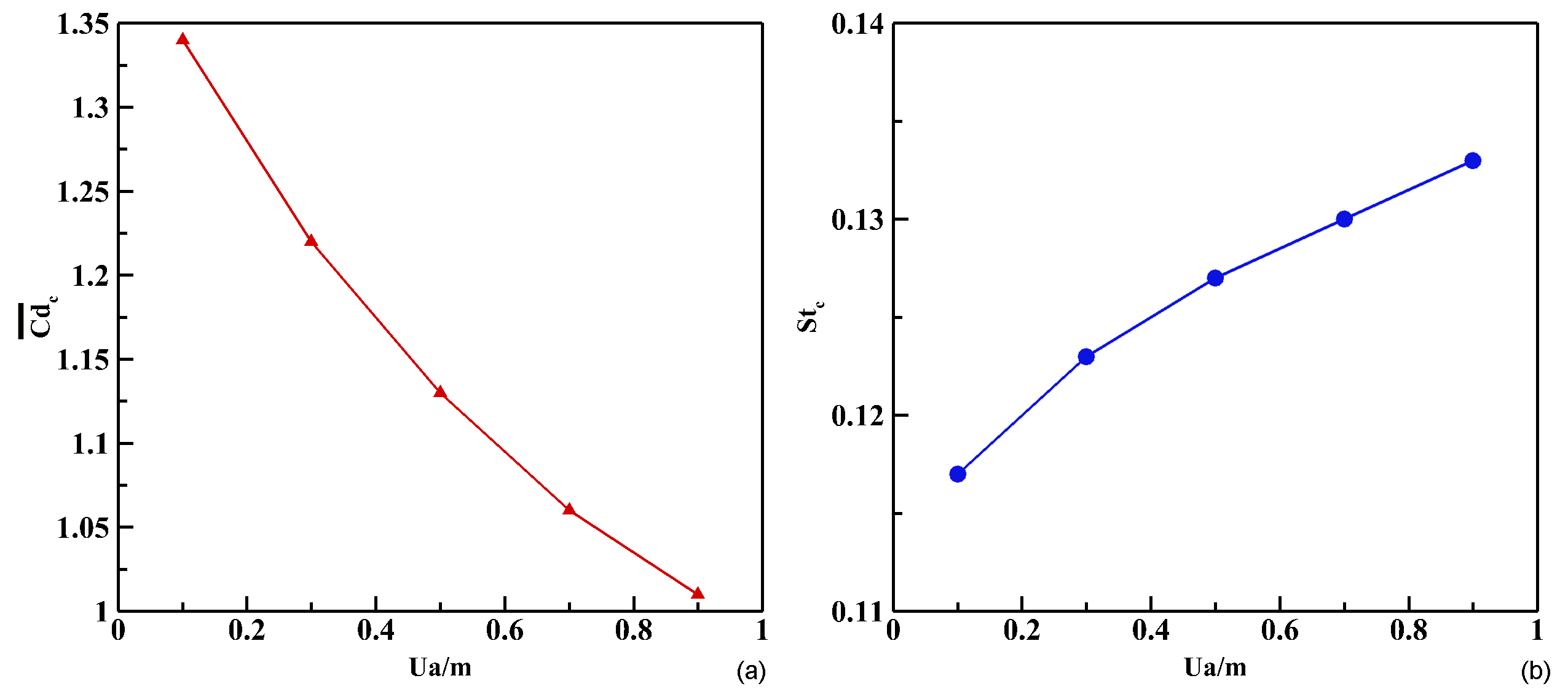}
  \caption{Relationship between (a) the critical drag coefficient $\overline{Cd}_c$ and (b) the critical Strouhal number $St_c$ with the parameter $Ua/m$. Red rectangular symbols represent $\overline{Cd}_c$, and blue circular symbols represent $St_c$ across varying $Ua/m$ values.}
\label{fig6}
\end{figure}

Building on these flow characteristics, the analysis extends to other dimensionless quantities at the critical Reynolds number, specifically the drag coefficient and Strouhal number.
Previous studies by~\citet{RASTAN2022} have shown a linear relationship between the critical Strouhal number $St_c$ and AR for rectangular cylinders, while those by~\citet{thompson2014} have demonstrated a nonlinear one for elliptical cylinders.
This prompts an investigation into similar trends for Rankine ovals.
As observed in Fig.~\ref{fig6}, Rankine ovals exhibit a nonlinear relationship akin to that of elliptical cylinders.

Figure~\ref{fig6} depicts the variation of the critical drag coefficient $\overline{Cd}_c$ and Strouhal number $St_c$ with $Ua/m$.
As $Ua/m$ increases, $\overline{Cd}_c$ decreases from 1.34 to 1.01, and $St_c$ rises from 0.12 to 0.13.
Notably, the product $\overline{Cd}_c \times St_c$ remains nearly constant at $0.14 \pm 0.01$, consistent with experimental findings by~\citet{Alam2008} for isolated bluff bodies.
This invariance across geometries, angles of attack, and Reynolds numbers underscores a robust characteristic in fluid dynamics, now verified for Rankine ovals.

\subsection{\label{subsec:Pressure Distributions from Potential Flow and Numerical Simulation}Pressure Distributions from Potential Flow and Numerical Simulation}
\begin{figure}
\centering
\includegraphics[width=1.0\textwidth]{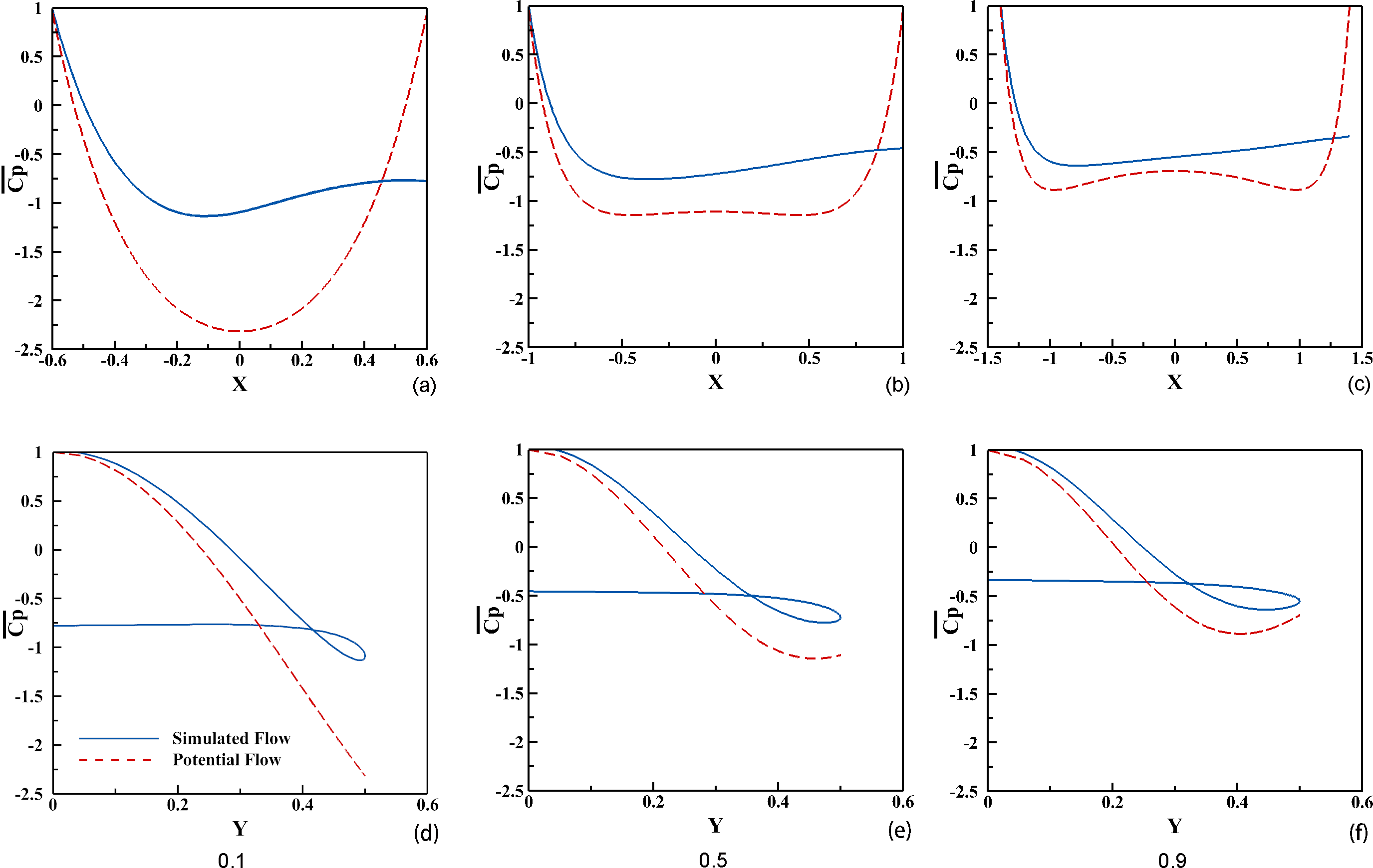}
\caption{Comparative analysis of the time-averaged pressure coefficient distributions on Rankine ovals at $Re=200$, contrasting results from numerical simulations with those from inviscid potential flow theory. The blue solid line represents the time-averaged pressure coefficient from the numerical simulation, while the red dashed line denotes the pressure coefficient from potential flow theory. Panels (a)-(c) illustrate distributions along the $x$-direction for $Ua/m = 0.1$, $0.5$, and $0.9$, respectively, while panels (d)-(f) show the corresponding distributions along the $y$-direction.}
\label{fig14}
\end{figure}

Figures~\ref{fig14}(a)-(c) illustrate the time-averaged surface pressure coefficient distributions along the $x$-axis of Rankine ovals for varying $Ua/m$.
In all cases, the pressure coefficient at the stagnation point attains its maximum value of 1.0.
As $Ua/m$ increases, the inviscid potential flow solution transitions from a single minimum to two distinct minima, accompanied by an increase in the minimum pressure coefficient, signifying a weaker suction peak.
In contrast, the numerical simulations, which account for viscous effects, consistently exhibit a single minimum that is generally less negative than the potential flow prediction.
This discrepancy arises primarily from viscous dissipation.
From the stagnation point, the pressure coefficient decreases along the surface, forming a minimum region, and then recovering at the rear of the body.
The minimum is located in the front half of the Rankine oval across all cases.
For $Ua/m < 0.5$, the simulated minimum is positioned closer to the leading edge than the central minimum in the potential flow solution.
However, for $Ua/m \geq 0.5$, this trend reverses, with the simulated minimum shifting rearward and approaching the potential flow location.
The deviation between the numerical and inviscid potential flow solutions diminishes as $Ua/m$ increases, reflecting reduced viscous influences at higher $Ua/m$.
This convergence suggests that, for sufficiently large $Ua/m$, the pressure drag coefficient may be approximated using the inviscid potential flow solution, potentially augmented by a correction for viscous effects.
Figures~\ref{fig14}(d)-(f) further examine this by presenting the time-averaged surface pressure coefficient distributions along the $y$-axis for different $Ua/m$ values.
Notably, the pressure coefficient in the rear half of the body approaches a plateau, indicative of a nearly uniform pressure region.
This observation indicates that the pressure distribution in the rear part of the body can be reasonably approximated by the pressure at the apex along the $y$-axis, while the front part is represented by the inviscid potential flow solution, as expressed by the piecewise approximation
\begin{equation}
C_{p, \text{visc}}(x) =
\begin{cases}
    C_{p, \text{pot}}(x), & \text{if } x < 0 \\
    C_{p, \text{pot}}(0), & \text{if } x \geq 0
\end{cases}
\label{eq:cp_approx}
\end{equation}
where $C_{p, \text{pot}}(x)$ is the inviscid potential flow solution, while $C_{p, \text{visc}}(x)$ is the estimated viscous pressure distribution.
For $x \geq 0$, as $x$ increases and $y$ correspondingly decreases, $C_p$ exhibits a plateau value.
This approximation is particularly apt for high-$Ua/m$ Rankine ovals, where pressure recovery in the wake is relatively flat.

To evaluate the accuracy of this approximation, the estimated pressure drag coefficient is compared with numerical results.
At $Ua/m = 0.1$, the estimation error exceeds 50\%, rendering the approach unreliable.
For $Ua/m = 0.5$, the error reduces to 32.5\%, suggesting marginal applicability.
At $Ua/m = 0.9$, the estimated pressure drag coefficient is $0.56$, within 5.6\% of the numerical value.
These findings demonstrate that, for sufficiently elongated Rankine ovals (i.e., large $Ua/m$), inviscid potential flow solutions, combined with a rear pressure correction, yield reasonably accurate estimates of pressure drag.

\subsection{\label{subsec:dragCoefficientsAndStrouhalNumbersAcrossReynoldsNumbersAndRankineGeometries}Drag Coefficient and Strouhal Number Across Different Reynolds Numbers and Shapes}
Building on the observed relationships between the drag coefficient, Strouhal number, and the $Ua/m$ parameter at the critical Reynolds number, we now extend the investigation to broader Reynolds number ranges for various Rankine oval shapes.

\begin{figure}
  \centering
      \includegraphics[width=1.0\textwidth]{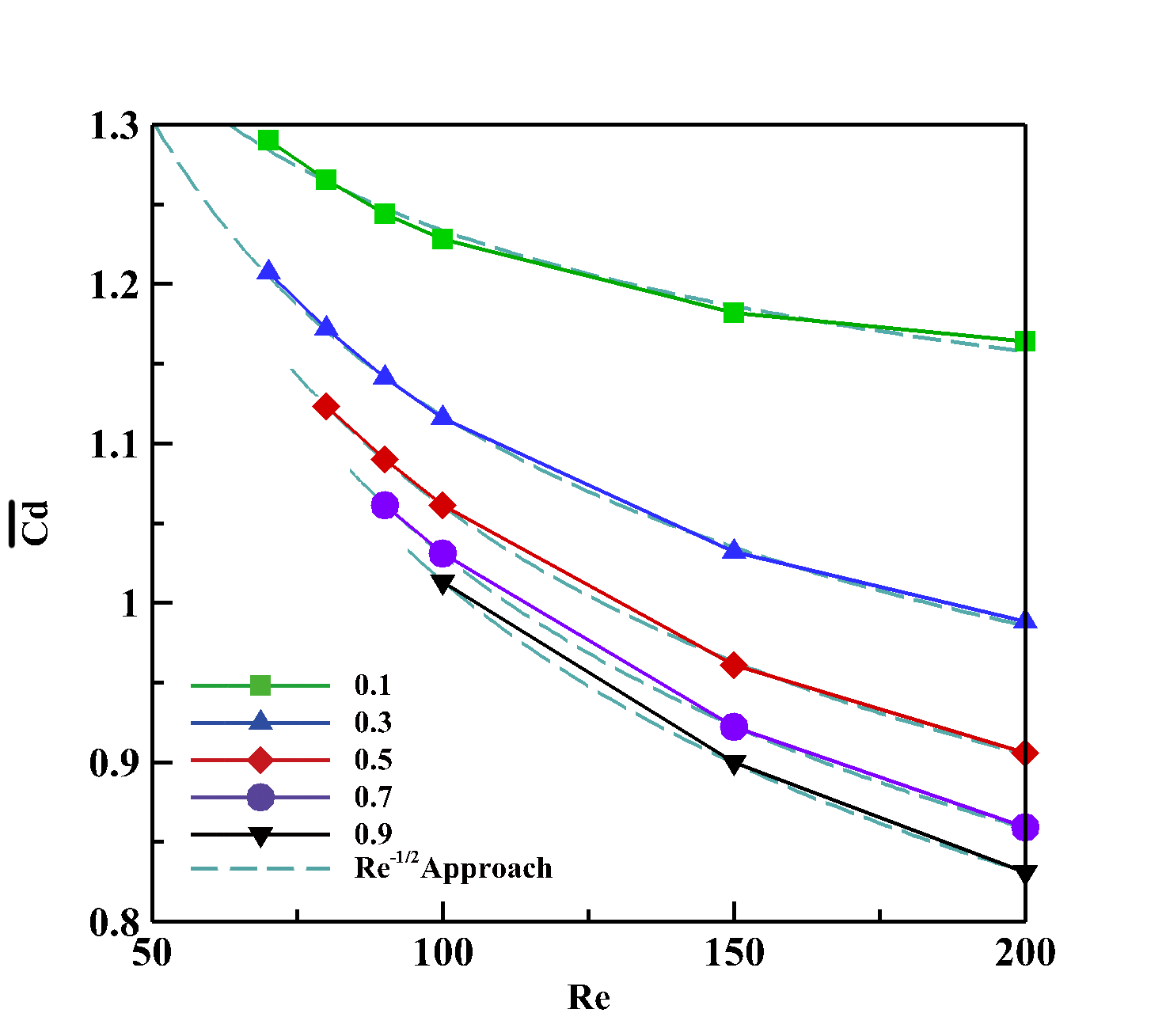}
  \caption{Relationship between the time-averaged drag coefficient $\overline{Cd}$ and Reynolds number $Re$ for various Rankine oval shapes. Symbols denote different $Ua/m$ values: green rectangles for $Ua/m=0.1$, blue triangles for $Ua/m=0.3$, red diamonds for $Ua/m=0.5$, purple circles for $Ua/m=0.7$, and black nablas for $Ua/m=0.9$. Dashed lines represent fitted models of the form $\overline{Cd} = a Re^{-1/2} + b$, where each line corresponds to specific coefficients $a$ and $b$, with $b \approx \overline{Cd}_p$. The trends show a decrease in $\overline{Cd}$ with increasing $Re$.}
\label{fig7}
\end{figure}

Figure~\ref{fig7} illustrates the relationship between the time-averaged drag coefficient $\overline{Cd}$ and Reynolds number $Re$ ranging from 50 to 200 for various Rankine oval shapes.
As $Re$ increases within this range, $\overline{Cd}$ decreases, though the rate of decrease slows.
For example, at $Ua/m=0.3$, $\overline{Cd}$ falls from 1.21 to 0.99.
This trend holds across all shapes examined.
Furthermore, for a fixed $Re$, $\overline{Cd}$ decreases with increasing $Ua/m$.
For instance, at $Re=100$, $\overline{Cd}$ reduces from 1.23 at $Ua/m=0.1$ to 1.03 at $Ua/m=0.9$, indicating lower drag for more elongated ovals.
These observations align with findings for rectangular and elliptical geometries, as reported by \citet{MASHHADI2021} and \citet{Raman2013}.

Numerical fitting yields the relation $\overline{Cd} = a Re^{-1/2} + b$, as shown by the dashed lines in Fig.~\ref{fig7}.
Each dashed line has unique $a$ and $b$ values, and $b \approx \overline{Cd}_p$.
Therefore, the relationship can be expressed as $\overline{Cd} = a Re^{-1/2} + \overline{Cd}_p$.
This implies that the friction drag component $\overline{Cd}_f$ scales linearly with $Re^{-1/2}$, which is inherently related to boundary layer theory.

\begin{figure}
  \centering
  \includegraphics[width=1.0\textwidth]{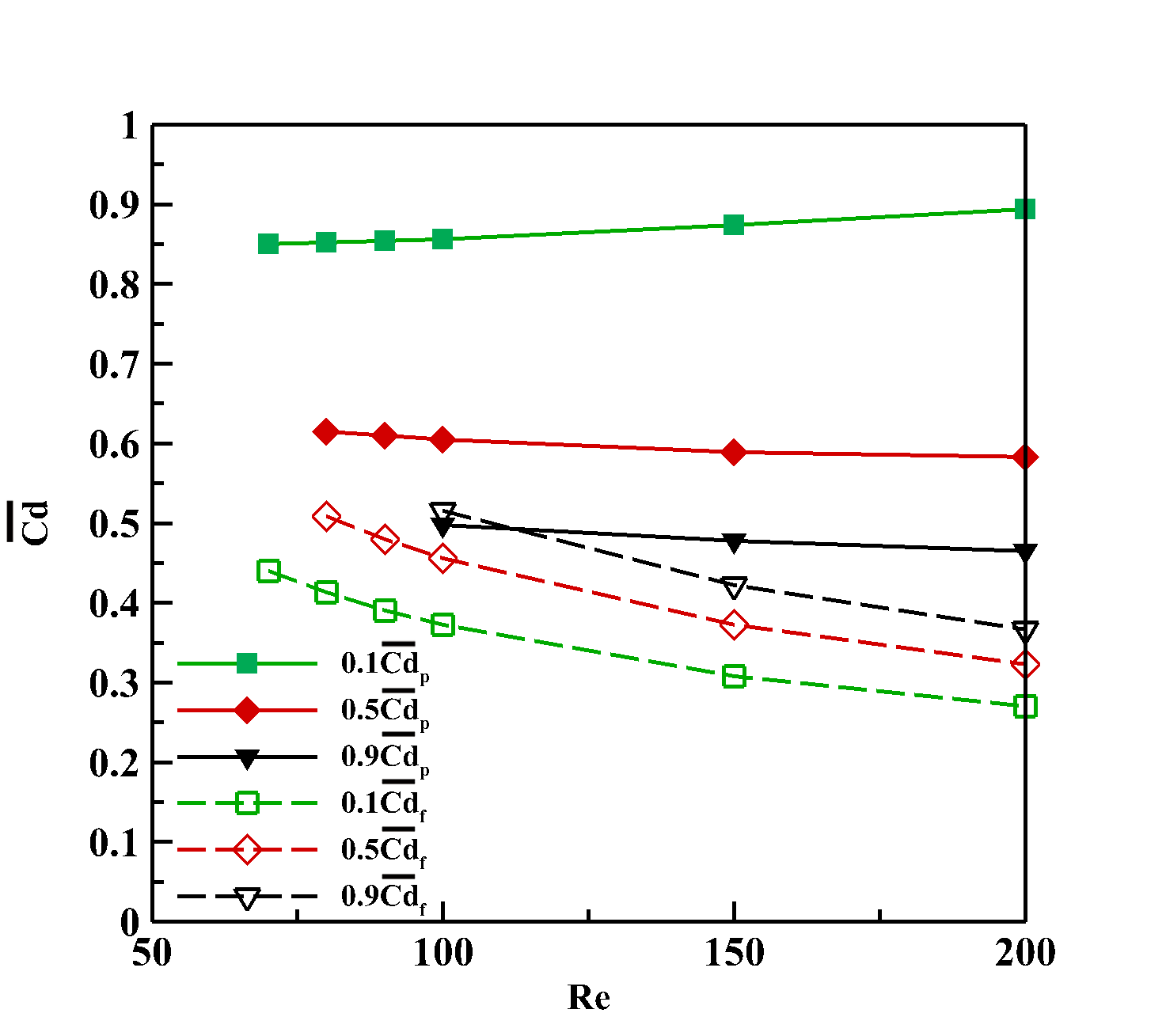}
  \caption{Variation of the time-averaged pressure drag coefficient $\overline{Cd}_p$ (solid lines) and friction drag coefficient $\overline{Cd}_f$ (dashed lines) with Reynolds number $Re$. Green rectangles denote $Ua/m=0.1$, red diamonds $Ua/m=0.5$, and black nablas $Ua/m=0.9$.}
\label{fig8}
\end{figure}
To gain deeper insight, we decompose the drag into its pressure and friction components, examining their contributions for different Rankine oval shapes across varying $Re$.
As shown in Fig.~\ref{fig8}, $\overline{Cd}_p$ remains nearly constant with increasing $Re$, while $\overline{Cd}_f$ decreases.
For $Ua/m=0.5$, $\overline{Cd}_p$ stays around 0.60 across all $Re$, whereas $\overline{Cd}_f$ drops from 0.52 to 0.35.
Notably, although $\overline{Cd}_f$ is typically lower than $\overline{Cd}_p$, it exceeds $\overline{Cd}_p$ at $Ua/m=0.9$ and $Re=100$, where $\overline{Cd}_f=0.53$ is larger than $\overline{Cd}_p=0.50$.

\begin{figure}
  \centering
  \includegraphics[width=1.0\textwidth]{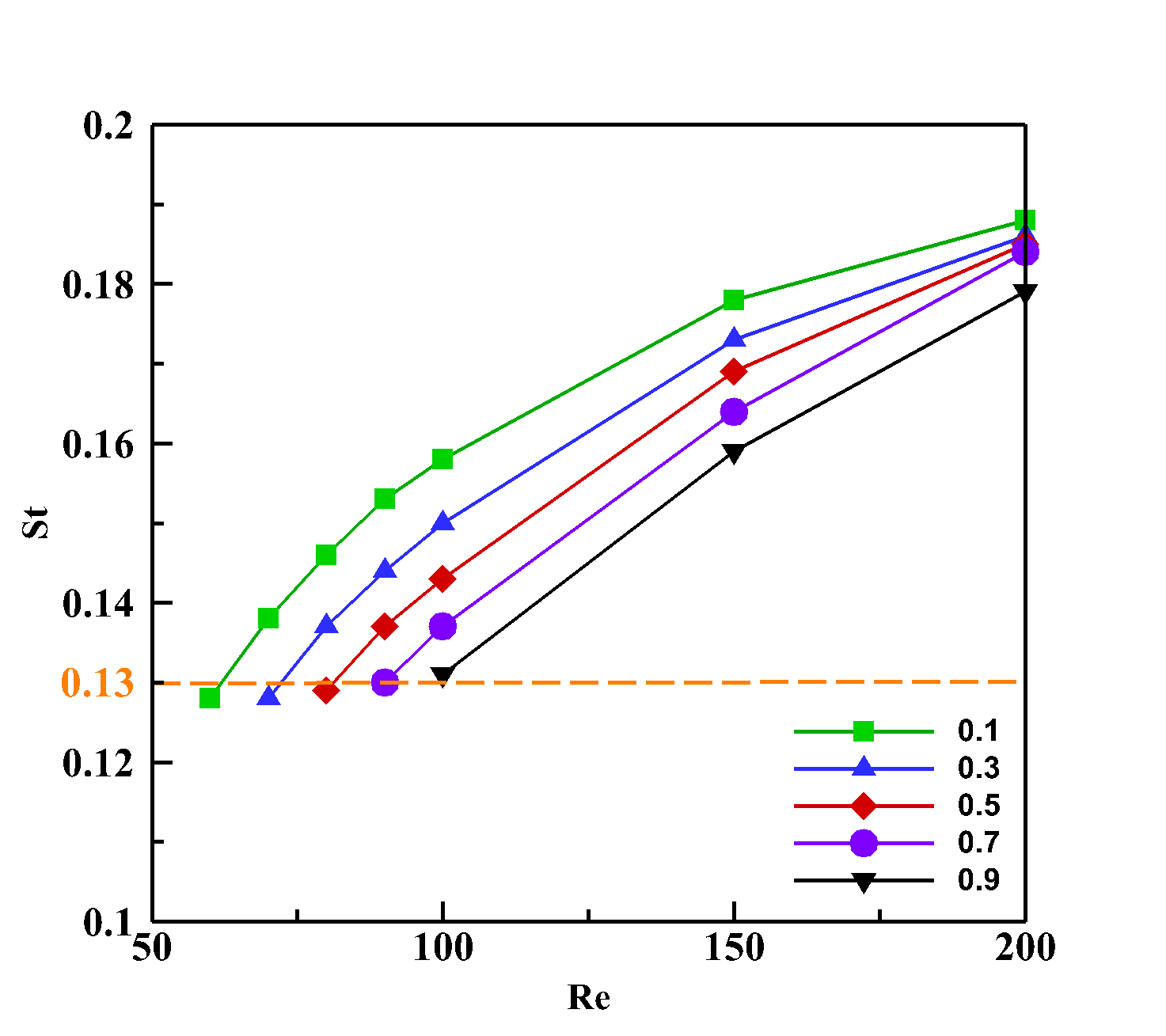}
  \caption{Relationship between the Strouhal number $St$ and Reynolds number $Re$ for various Rankine oval shapes. Symbols and colors indicate different $Ua/m$ values: green rectangles for $Ua/m=0.1$, blue triangles for $Ua/m=0.3$, red diamonds for $Ua/m=0.5$, purple circles for $Ua/m=0.7$, and black nablas for $Ua/m=0.9$. The orange dashed line corresponds to $St=0.13$, highlighting the trend above the critical Reynolds number.}
\label{fig9}
\end{figure}
Shifting focus to vortex shedding dynamics, we next explore the relationship between $St$ and $Re$ for different Rankine oval configurations.
Figure~\ref{fig9} reveals that, for a given shape, $St$ increases with $Re$, reflecting higher vortex shedding frequencies at elevated flow velocities.
For example, at $Ua/m=0.1$, $St$ rises from 0.13 at $Re=60$ to 0.19 at $Re=200$.
At $Re=200$, $St$ remains approximately 0.18 for all values of $Ua/m$, suggesting that vortex shedding frequencies become largely independent of geometries in this high-$Re$ regime.
Additionally, increasing $Ua/m$, corresponding to more elongated Rankine ovals, which in turn reduces $St$.
For instance, at $Re=100$, $St$ decreases from 0.16 at $Ua/m=0.1$ to 0.13 at $Ua/m=0.9$.
Moreover, as indicated by the orange dashed line, when $Re$ exceeds the critical value $Re_c$, $St$ stabilizes at approximately 0.13 rather than increasing gradually from zero.

\subsection{\label{subsec:vortexSheddingStructure}Vortex Shedding Structure}
As $St$ varies with $Re$ and $Ua/m$, we further investigate the flow field across different $Re$ and $Ua/m$ values, focusing on the corresponding changes in vortex shedding structure.
\begin{figure}
  \centering
  \includegraphics[width=1.0\textwidth]{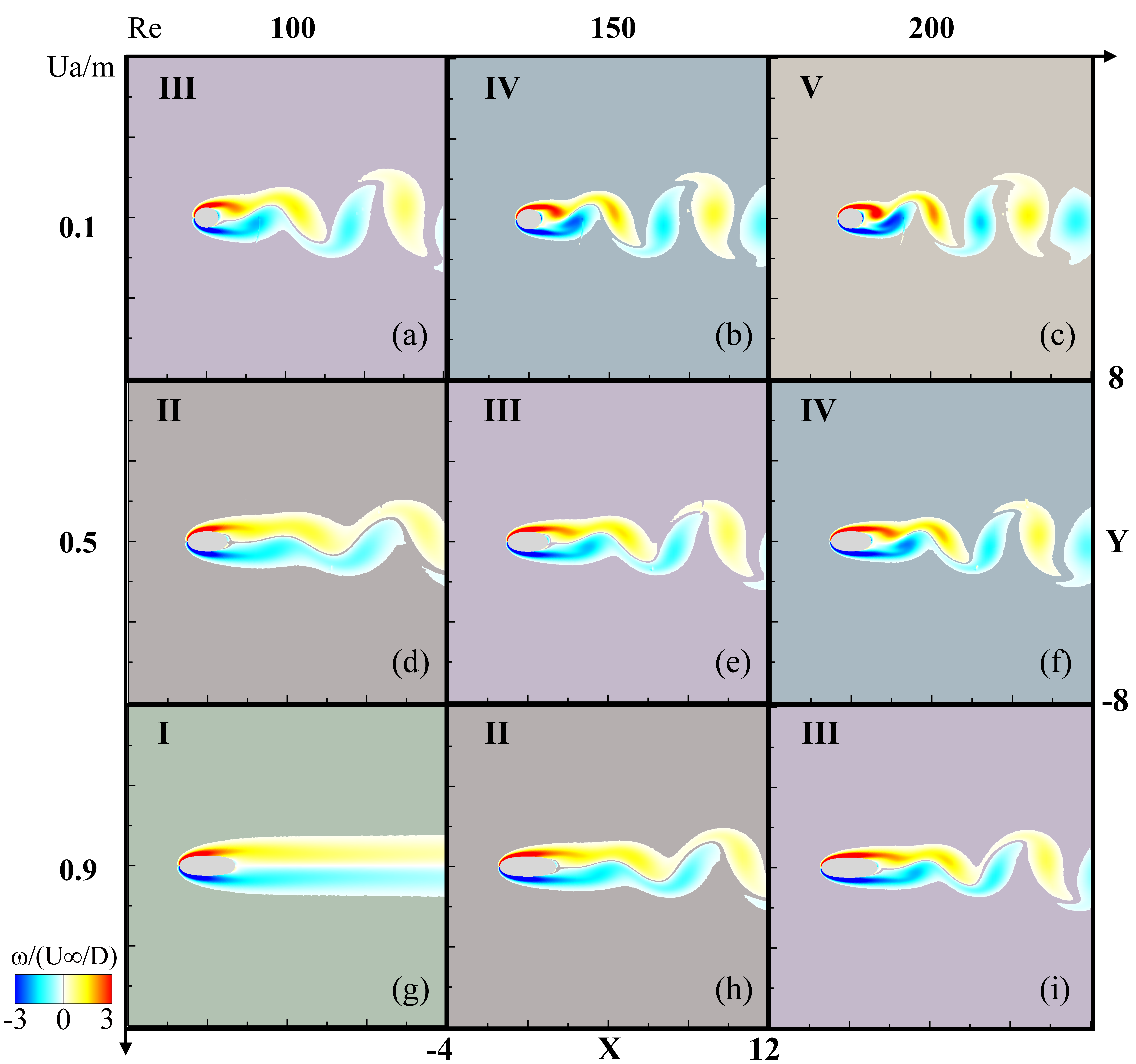}
  \caption{
Instantaneous vorticity fields around Rankine ovals at Reynolds numbers of 100, 150, and 200, and $Ua/m$ values of 0.1, 0.5, and 0.9. The subfigures are divided into five distinct groups (I, II, III, IV, and V) based on wake vortex characteristics, with each group denoted by a unique background color. Vorticity is colored from blue to red, spanning values from $-3$ to $3$. The Rankine oval (grey) is located at the origin $(0,0)$. The flow field is depicted within the two-dimensional Cartesian range $x \in [-4,12]$ and $y \in [-8,8]$.
}
\label{fig10}
\end{figure}
In Fig.~\ref{fig10}, the instantaneous flow fields around Rankine ovals are depicted under conditions with varying Reynolds numbers and $Ua/m$ values.
A key observation is the increase in both vorticity magnitude and shedding frequency in the wake as Reynolds number rises and $Ua/m$ decreases.
This trend is evident when comparing subfigures from the bottom left, which correspond to lower $Re$ and higher $Ua/m$, to the top right, which represent higher $Re$ and lower $Ua/m$.
We organize the subfigures into five distinct groups (I to V) with background colors based on similarities in wake vortex intensity and frequency.
The grouping follows a diagonal pattern from bottom right to top left: Group I contains only subfigure (g); Group II includes (d) and (h); Group III comprises (a), (e), and (i) along the central diagonal; Group IV consists of (b) and (f); and Group V is solely subfigure (c) at the top right.
This grouping effectively demonstrates the combined effects of $Re$ and $Ua/m$ on vortex shedding structures, as it organizes subfigures based on similarities in wake vortex intensity and frequency, thereby highlighting the systematic variations in wake dynamics driven by their joint influence and suggesting the potential for normalization through a unified parameter.
 
\subsection{\label{subsec:velocityFluctuations}Velocity Fluctuations}
\begin{figure}
  \centering
  \includegraphics[width=1.0\textwidth]{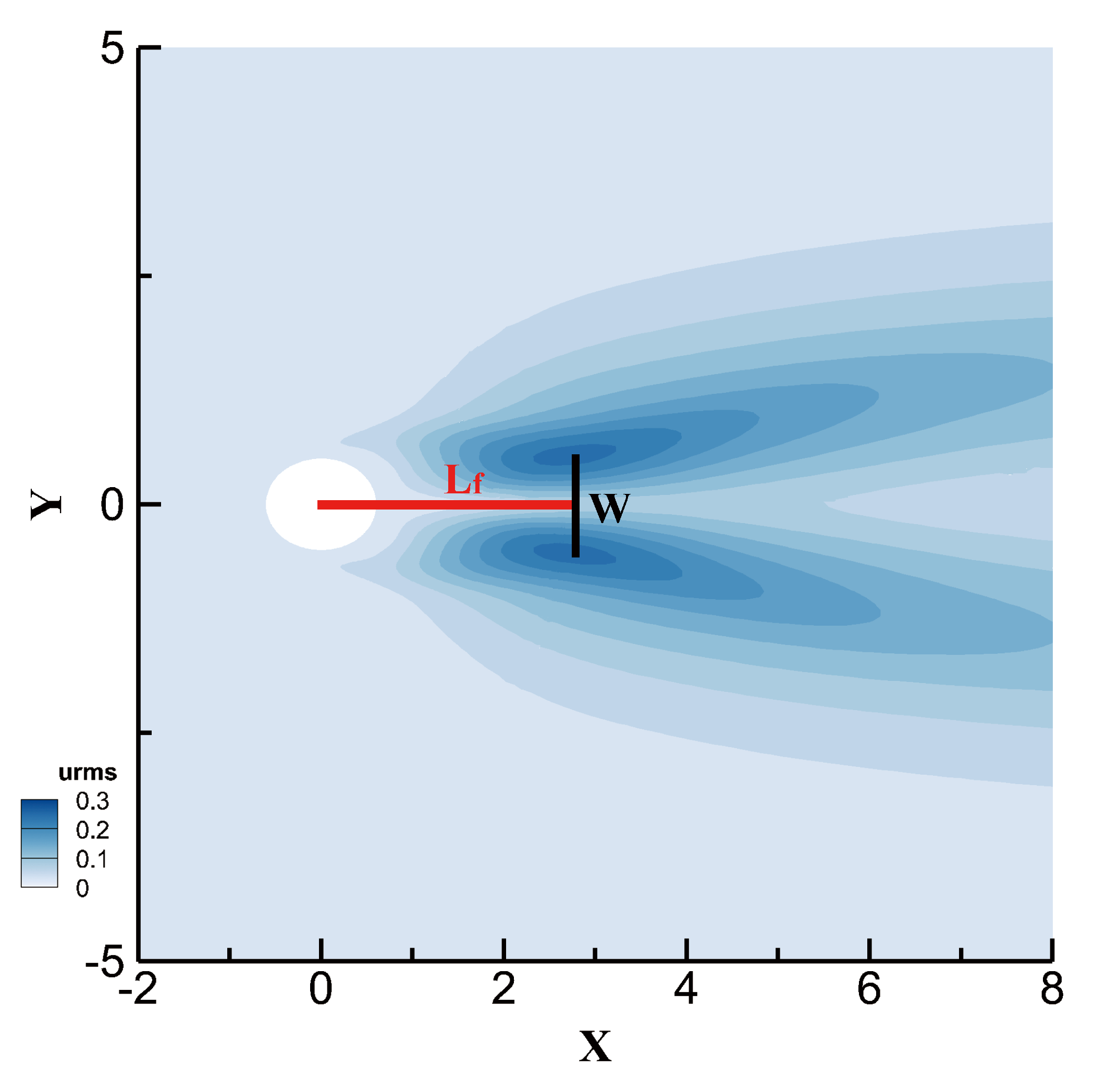}
  \caption{The root-mean-square streamwise fluctuating velocity in the flow around a Rankine oval at $Re=100$ and $Ua/m=0.1$. It highlights the vortex formation length $L_f$ (red line) and the wake width $W$ (black line).}
\label{fig11}
\end{figure}
The root-mean-square (rms) fluctuating velocity is a key feature of the flow.
\citet{Gerrard_1966} introduced the concepts of wake width $W$ and vortex formation length $L_f$ in bluff body research.
As shown in Fig.~\ref{fig11} and based on the definition of \citet{RASTAN2022}, the wake width $W$ is defined as the distance between the two distinct peaks of the rms fluctuating velocity, which is marked by a solid black line.
The vortex formation length $L_f$ is defined as the streamwise distance from the center of the Rankine oval to the midpoint between these two peaks, and it is denoted by a red line.
Thus, $L_f$ represents the extent of the vortex formation region in the flow field.

\begin{figure}
  \centering
  \includegraphics[width=1.0\textwidth]{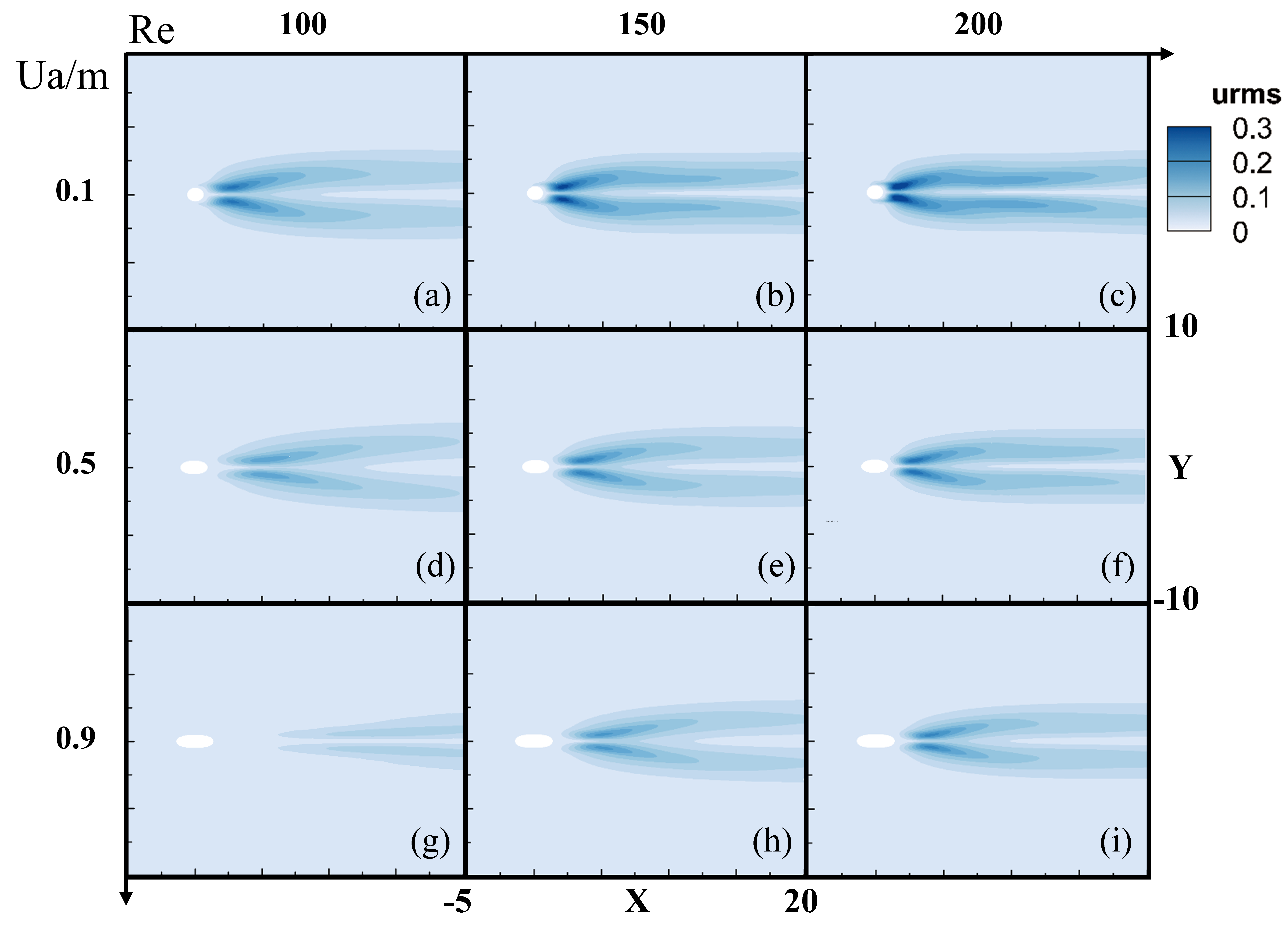}
\caption{The rms velocity distributions for various Rankine ovals at Reynolds numbers of 100, 150, and 200, and $Ua/m$ values of 0.1, 0.5, and 0.9. The color bar ranges from white to blue, indicating rms streamwise velocity from 0 to 0.3.}
\label{fig12}
\end{figure}
We analyze the rms velocity distributions, as illustrated in Fig.\ref{fig12}, and compute the corresponding wake widths and vortex formation lengths.
In Fig.\ref{fig12}, the Rankine oval is positioned at the origin $(0,0)$, with the flow field visualized over $x \in [-5,20]$ and $y \in [-10,10]$.
As Reynolds number increases and $Ua/m$ decreases, the maximum rms velocity rises significantly, and the rms peaks shift closer to the Rankine oval.
These observations indicate that the flow around Rankine ovals is strongly influenced by both $Re$ and $Ua/m$.

\begin{table}
\centering
\caption{\label{LfAndW}Vortex formation length $L_f$ and wake width $W$ for different Reynolds numbers and Rankine ovals.}
\begin{tabular}{c|cc|cc|cc}
\toprule
\multirow{2}{*}{\raisebox{-0.5ex}{$Ua/m$}} & \multicolumn{2}{c|}{Re=100} & \multicolumn{2}{c|}{Re=150} & \multicolumn{2}{c}{Re=200} \\

\cmidrule{2-7}
                        & $L_f$ & $W$ & $L_f$ & $W$ & $L_f$ & $W$ \\
\midrule
0.1                     & 2.73 & 1.02 & 2.06 & 0.93 & 1.72 & 0.85 \\
0.3                     & 4.01 & 1.12 & 2.85 & 0.94 & 2.33 & 0.85 \\
0.5                     & 5.27 & 1.13 & 3.58 & 0.97 & 2.96 & 0.86 \\
0.7                     & 7.38 & 1.18 & 4.35 & 0.97 & 3.49 & 0.86 \\
0.9                     & 15.62 & 1.41 & 5.10 & 1.01 & 4.05 & 0.90 \\
\bottomrule
\end{tabular}
\end{table}

Table~\ref{LfAndW} presents the vortex formation length $L_f$ and wake width $W$ for five Rankine ovals across different Reynolds numbers.
At a fixed $Re$, larger $Ua/m$ values lead to increases in both $W$ and $L_f$, indicating expanded distances between rms peaks and longer vortex formation regions.
For instance, at $Re=100$, $L_f$ is markedly larger for $Ua/m=0.9$ with a value of 15.62 compared to $Ua/m=0.1$ with a value of 2.73.
For a fixed $Ua/m$, increasing $Re$ reduces both $W$ and $L_f$, with greater reductions observed for larger $Ua/m$ values.
These findings are consistent with prior studies on flow separation around two-dimensional bodies~\citep{RASTAN2022}.
Notably, at $Re=200$, the wake width $W$ approaches a nearly constant value of approximately 0.86 for most shapes, suggesting that wake width becomes substantially independent of $Ua/m$ in this regime.

\subsection{\label{subsec:instantaneousLiftAndDragCoefficients}Instantaneous Lift and Drag Coefficient}
While previous sections focused on mean flow quantities, we investigate the instantaneous coefficients to reveal the unsteady fluid dynamics, particularly for lift coefficient, whose average value theoretically remains zero despite significant instantaneous variations.
\begin{figure}
  \centering
  \includegraphics[width=1.0\textwidth]{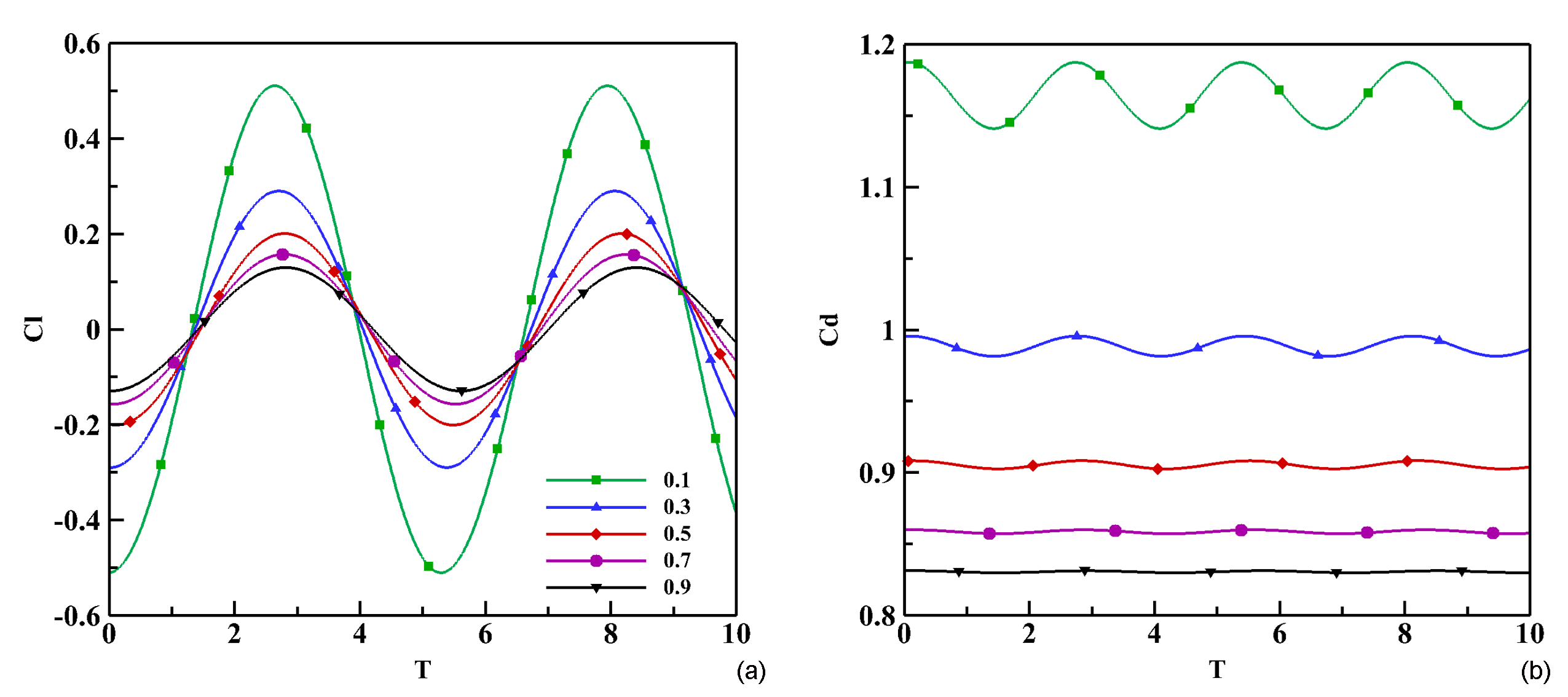}
    \caption{Instantaneous lift and drag coefficients across various Rankine ovals at $Re = 200$. Instantaneous coefficients are depicted as lines, distinguished by colors and symbols: green rectangles for $Ua/m = 0.1$, blue triangles for $Ua/m = 0.3$, red diamonds for $Ua/m = 0.5$, purple circles for $Ua/m = 0.7$, and black inverted triangles for $Ua/m = 0.9$.}
    \label{fig16}
    \end{figure}
In Fig.~\ref{fig16}, we observe variations in the amplitude of $Cl$ and $Cd$. 
For clarity, we choose to present only five distinct $Ua/m$ values.
We adjusted the phase of the coefficients in our figures for enhanced comparative clarity.
The amplitude of the lift and drag coefficients decreases as $Ua/m$ increases.
The maximum instantaneous lift coefficient for $Ua/m=0.1$ is $Cl=0.51$, which diminishes to $Cl=0.13$ for $Ua/m=0.9$, while the average lift coefficient remains consistently at zero.
Similarly, the maximum and average instantaneous drag coefficients for $Ua/m=0.1$ are $Cd=1.19$ and $\overline{Cd}=1.16$, respectively, decreasing to $Cd=0.83$ and $\overline{Cd}=0.83$ for $Ua/m=0.9$.
An inverse relationship is observed between $Ua/m$ and the frequency of these coefficients, aligning with the Strouhal number trends observed in Fig.~\ref{fig9}. 
To comprehensively understand the amplitude trends across all $Ua/m$ values, we further exhibit the amplitudes of both $Cl$ and $Cd$ in Fig.~\ref{fig17}.

\begin{figure}
  \centering
  \includegraphics[width=1.0\textwidth]{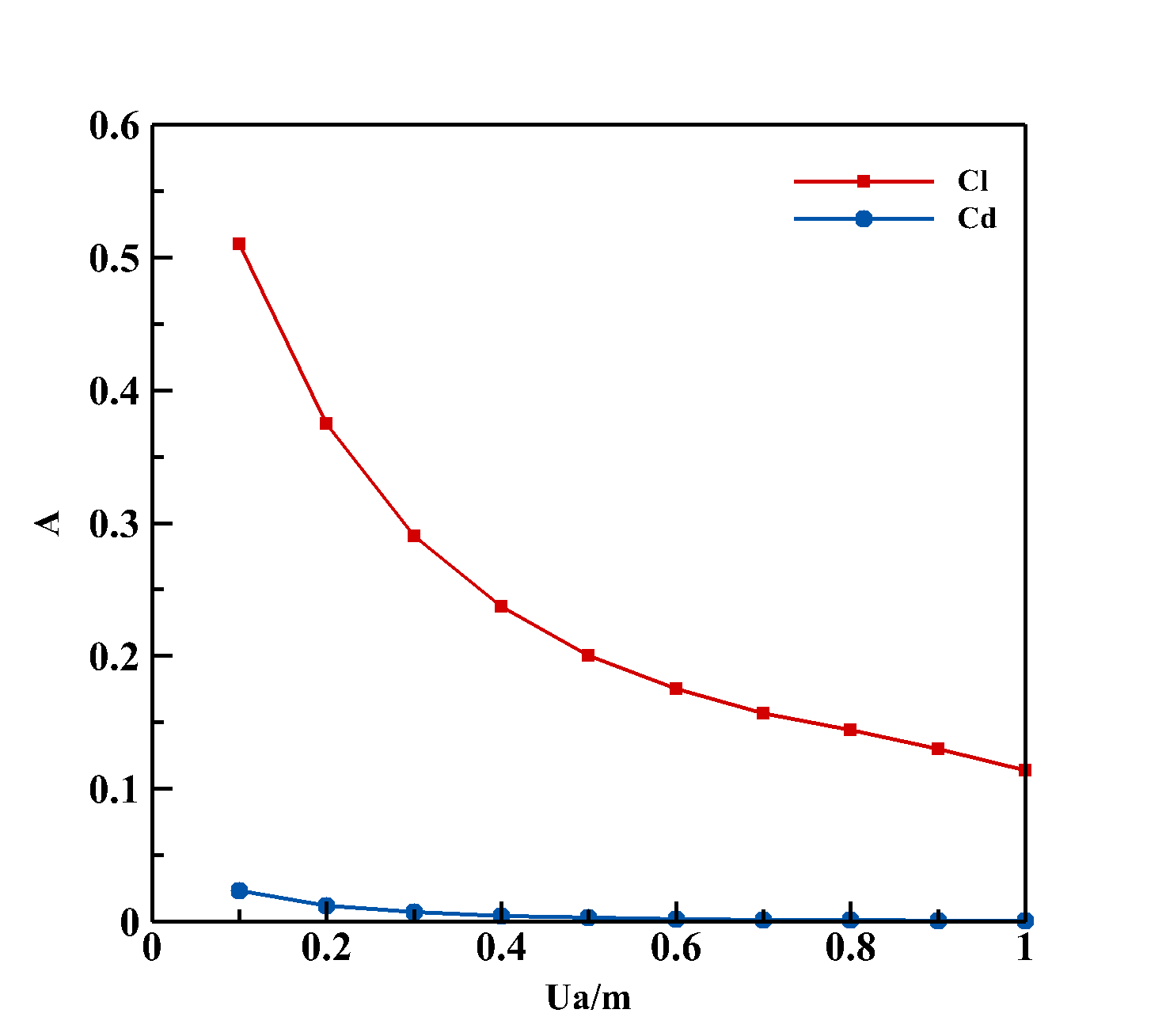}
\caption{The amplitude of $Cd$ and $Cl$ at $Re=200$ for various $Ua/m$. Blue circles represent the cases of $Cd$, while red rectangles depict those of $Cl$.}
\label{fig17}
\end{figure}
Figure~\ref{fig17} highlights the amplitude variations of lift and drag coefficients for various Rankine ovals at $Re = 200$.
The amplitude of $Cl$ is notably higher than that of $Cd$.
A decrease in $Cl$ amplitude, from 0.51 to 0.114, is observed with increasing $Ua/m$.
The $Cd$ amplitude exhibits a moderate decreasing trend, changing from 0.0235 to 0.0004.
These results suggest that elongated Rankine ovals suppress the amplitudes of the lift and drag coefficients by mitigating the intensity of vortex shedding and stabilizing the wake structure, despite the persistence of unsteady flow fluctuations.

Building upon this understanding, we further dissect the components of $Cd$ to discern the respective contributions of friction and pressure to the overall drag force.
This analysis aims to provide a comprehensive understanding of how these individual components influence the instantaneous drag coefficients.
\begin{figure}
  \centering
  \includegraphics[width=1.0\textwidth]{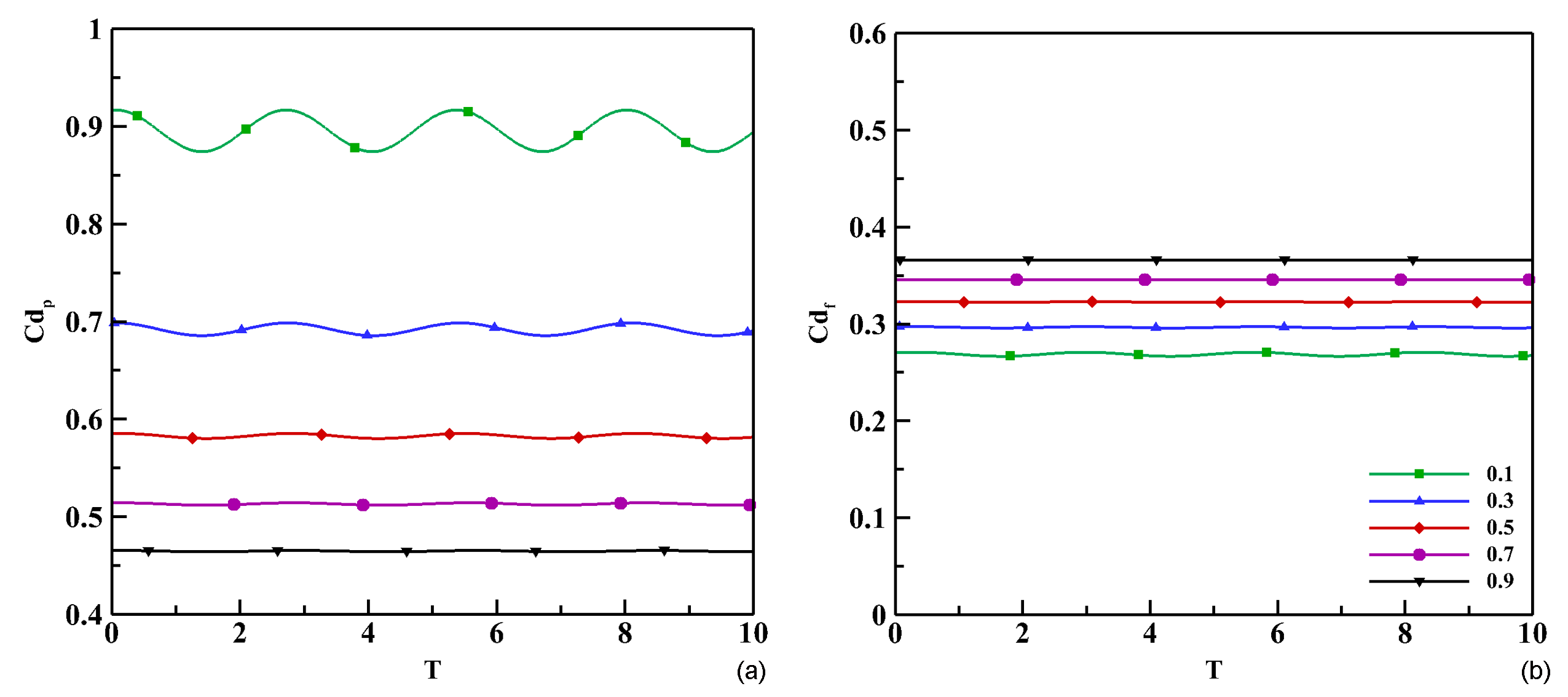}
\caption{The two main components of the instantaneous drag coefficient, pressure drag coefficient ($Cd_p$) and friction drag coefficient ($Cd_f$), for different Rankine ovals at $Re = 200$. The coefficients are depicted as lines, each distinguished by symbols and colors: green rectangles represent $Ua/m = 0.1$, blue triangles for $Ua/m = 0.3$, red diamonds indicate $Ua/m = 0.5$, purple circles for $Ua/m = 0.7$, and black inverted triangles correspond to $Ua/m = 0.9$.}
\label{fig18}
\end{figure}

As presented in Fig.~\ref{fig18}, the two components of the instantaneous drag coefficient are the pressure drag coefficient ($Cd_p$) and the friction drag coefficient ($Cd_f$).
There is a significant periodic fluctuation in $Cd_p$, while $Cd_f$ remains relatively constant.
For instance, at $Ua/m = 0.3$, the amplitudes of $Cd_p$ and $Cd_f$ are $0.007$ and $0.001$, respectively.
This pattern suggests that the periodic fluctuation in the drag coefficient is predominantly attributed to fluctuations in the pressure drag component.
Furthermore, as the length of the Rankine oval increases along the inlet flow direction, corresponding to a rise in $Ua/m$, there is a notable decrease in the amplitude of periodic fluctuation in the pressure drag component.
This reduction ranges from 0.042 at $Ua/m = 0.1$ to 0.001 at $Ua/m = 0.9$.
The periodic fluctuation of Rankine ovals with $Ua/m > 0.8$ become nearly indiscernible, attributed to the minimal amplitude of $Cd_p$.
Additionally, an increase in $Ua/m$ leads to an increase in the friction drag coefficient and a decrease in the pressure drag coefficient.

In summary, the analysis of instantaneous lift and drag coefficients reveals that increasing $Ua/m$ in Rankine ovals leads to suppressed amplitudes and reduced vortex shedding frequency.
The decomposition of drag into pressure and friction components further demonstrates that pressure drag dominates the unsteady fluctuations, with friction drag increasing for more elongated shapes.

\subsection{\label{subsec:dimensionanalysis}Data-driven Dimensional Analysis}
The Rankine oval flow involves numerous dimensional variables, including the half length $l$, half height $h$, vortex formation length $L_f$, wake width $W$, half-distance between the source and sink pair $a$, source-sink strength $m$, freestream flow speed $U$, largest reverse-flow speed $U_r$, frequency $f$, fluid density $\rho$ and viscosity $\mu$.
As discussed in our previous work~\citep{XU2022}, data-driven dimensional analysis is instrumental in identifying the dominant dimensionless quantities that govern the flow characteristics.
We employ the data-driven dimensional analysis method to simplify the complexity of the physical problem, which in this manuscript refers to the flow around Rankine ovals, by reducing the number of dimensionless quantities.

For the analysis of the frictional drag coefficient, the physical quantities selected are $\rho, \mu, l, a, U, m$, where $l$ is the half-length of Rankine oval. 
So the dependent dimensionless quantity $\overline{Cd}_{f}$ becomes $\overline{Cd}_{fl}={F_d^{fric}}/{0.5\rho U_\infty^2l}$.
And according to classical dimensional analysis, the dimensionless equation can be expressed as
\begin{equation}
\label{Cdrelation}
\overline{Cd}_{fl}=f \left(\frac{\rho{l}{U}}{\mu},\frac{Ua}{m},\frac{l}{a} \right).
\end{equation}
The corresponding dimensionless matrices $\mathbf{D}$, $\mathbf{W}^*$ and $\mathbf{W}$ for this problem are given by
\begin{equation}
\mathbf{D} = \begin{array}{*{20}{c}}
  {} \\ 
  M \\ 
  L \\ 
  T 
\end{array}\begin{array}{*{20}{c}}
  {\begin{array}{*{20}{c}}
  \rho &\mu & l &a & U & m
\end{array}} \\ 
  {\left[ {\begin{array}{*{20}{c}}
  1&1&0&0&0&0\\ 
  { - 3}&{ - 1}&1&1&1&2 \\ 
  0&{ - 1}&0&0&{ - 1}&{ - 1} 
\end{array}} \right]} 
\end{array}
\end{equation}
\begin{equation}
\mathbf{W}^* = \left[ \begin{array}{ccc}
1 & 0 & 0 \\
-1 & 0 & 0 \\
1 & 1 & 0 \\
0 & -1 & 1 \\
1 & 0 & 1 \\
0 & 0 & -1
\end{array} \right], \quad
\mathbf{W} = \left[ \begin{array}{ccc}
0.5 & -0.189 & -0.314 \\
-0.5 & 0.189 & 0.314 \\
0.5 & 0.567 & 0.210 \\
0 & -0.756 & 0.210 \\
0.5 & -0.189 & 0.420 \\
0 & 0 & -0.734
\end{array} \right].
\end{equation}

\begin{figure}
  \centering
  \includegraphics[width=1.0\textwidth]{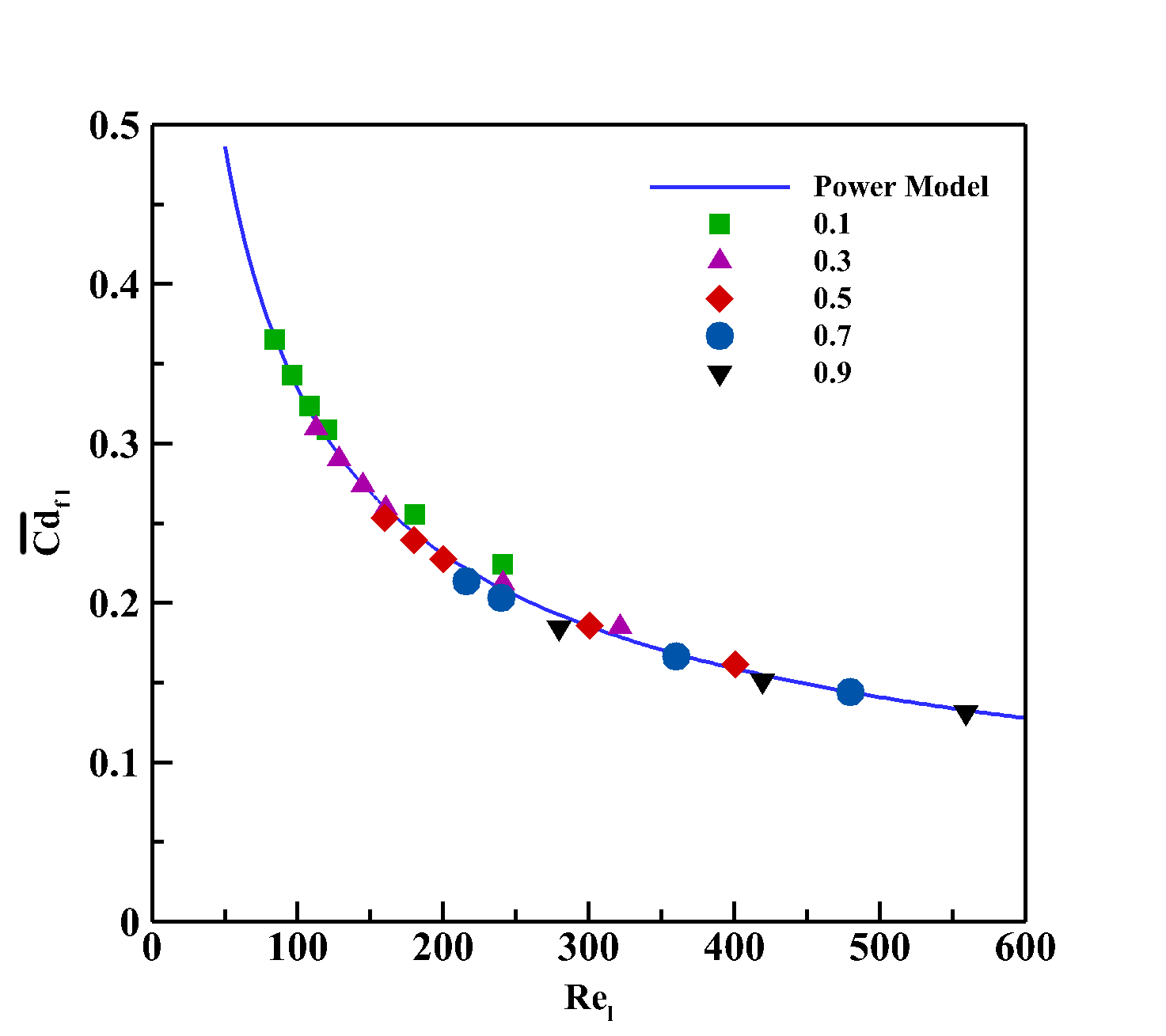}
\caption{Variation of the frictional drag coefficient $\overline{Cd}_{fl}$ as a function of the Reynolds number $Re_l$, with dimensionless quantities based on the length $l$.}
\label{fig19}
\end{figure}

The computational steps follow those outlined in~\citet{XU2022}.
Through the method, we identify the primary dimensionless quantity as
\begin{equation}
\label{keydimensionlessnumber}
\begin{split}
\overline{Cd}_{fl}=&g^*(\hat\pi_1)\\
=&g^*(\rho^{0.25}\mu^{-0.25}l^{0.25}a^{0.00}U^{0.25}m^{0.00}).
\end{split}
\end{equation}
The underlying principle of this method is to determine the most significant unit vector direction on the response surface, ensuring that the sum of the squared exponents in the dimensionless quantities equals 1.
The scaling of the dominant dimensionless quantity in~\eqref{keydimensionlessnumber} corresponds to the Reynolds number based on $l$, given by $Re_l = \rho l U / \mu$.
Thus, according to the data-driven dimensional analysis, the unique and important dimensionless quantity that determines the friction drag coefficient $\overline{Cd}_{fl}$ is $Re_l$.
As illustrated in Fig.~\ref{fig19}, the relationship between $\overline{Cd}_{fl}$ and $Re_l$ is independent of $U a/m$.
The line in Fig.~\ref{fig19} indicates that the frictional drag on Rankine ovals is predominantly influenced by the length $l$, which is aligned parallel to the freestream flow direction.
Regardless of other parameters, $l$ emerges as the primary determinant of frictional drag, as variations in it significantly alter the surface area exposed to the fluid flow along this direction, thereby directly affecting the frictional forces.

We fitted the data using a power-law model,
\begin{equation}
    \overline{Cd}_{fl} = 4.0 Re_l^{-0.54},
\end{equation}
which provides high precision with a Root Mean Square Error (RMSE) of 0.006. 
RMSE measures the differences between values predicted by the function and the values actually observed.
When combined with prior estimates of pressure drag coefficients with Eq.~\eqref{eq:cp_approx}, this suggests that the total drag for Rankine ovals with sufficiently large $Ua/m$ can be predicted accurately without numerical simulations. 
Specifically, for sufficiently large $Ua/m$ and a given $Re$, the surface pressure distribution is obtained through potential flow theory, which, when combined with Eq.~\eqref{eq:cp_approx} and numerical integration along axis, determines $Cd_p$.
Subsequently, $l$ is calculated based on $Ua/m$, allowing $Cd_{fl}$ to be derived from the power-law model using $Re_l$, converted from $Cd_{fl}$ to $ Cd_f $, and then added to $Cd_p$ to yield the total drag coefficient $Cd$.

\begin{figure}
  \centering
  \includegraphics[width=1.0\textwidth]{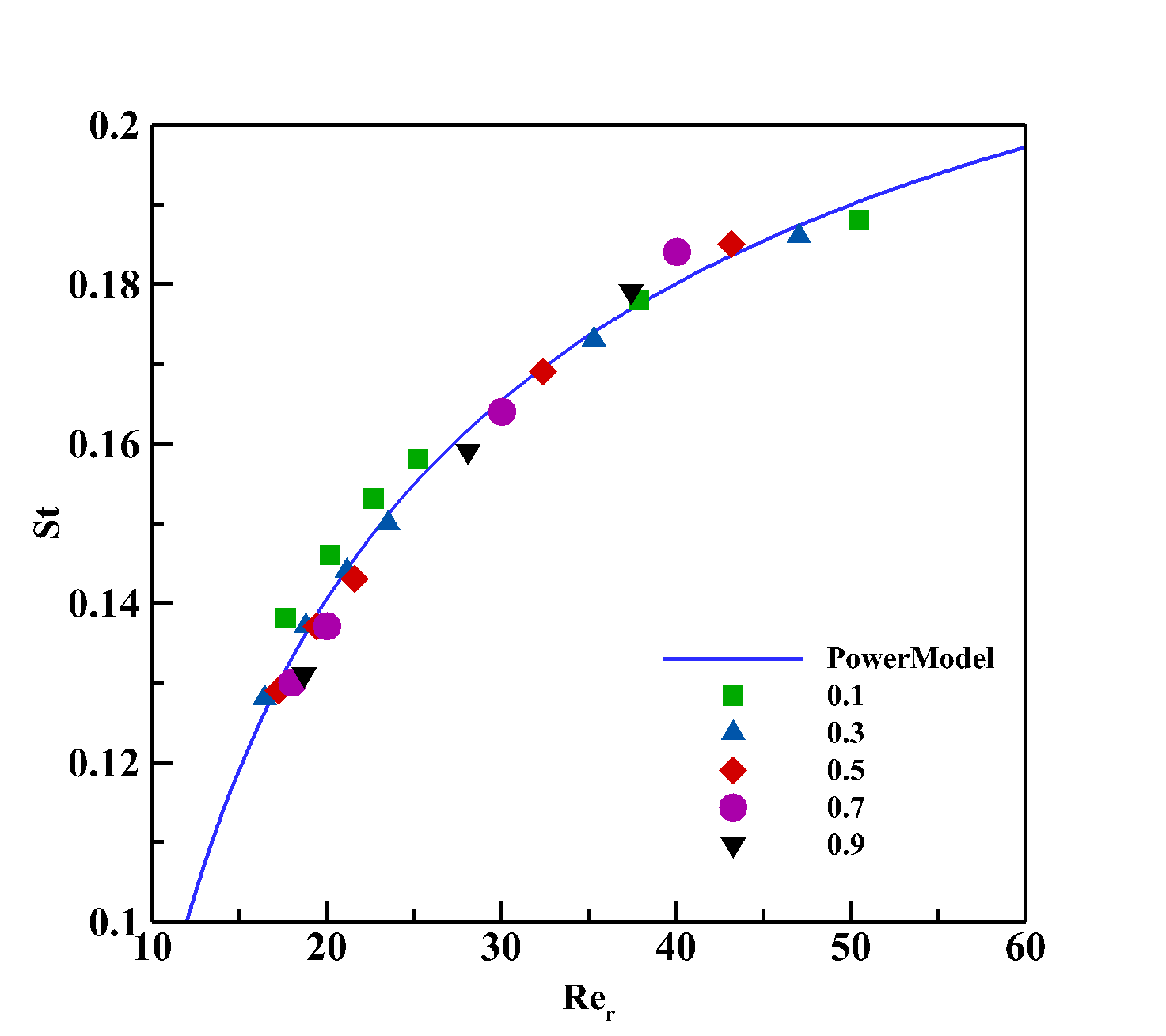}
\caption{Relationship between the Strouhal number $St$ and $Re_r$ for Rankine ovals with varying $Ua/m$, nondimensionalized using the vortex formation length $L_f$ and the largest reverse-flow speed $U_r$.}
\label{fig20}
\end{figure}
In our analysis of various result combinations, we further identified a dominant non-dimensional number, which is the Strouhal number $St$.
For the analysis of $St$, the physical quantities selected are $\rho, \mu, L_f, a, U_r, m$, where the velocity is the largest reverse-flow speed $U_r$, and the length is set as $L_f$.
Following data-driven dimensional analysis, $St$ is found to be independent of $Ua/m$ and dominated by the nondimensional parameter $Re_r = \rho U_r L_f / \mu$.
The relationship is fitted with a power function, $St = -0.54 Re_r^{0.54} + 0.27$, yielding an RMSE of 0.003, which confirms the high accuracy of the fit.

Figure~\ref{fig20} further demonstrates that $St$ is independent of $Ua/m$ and depends only on $Re_r$. 
This implies that the vortex shedding frequency is governed by the wake's physical characteristics, such as the largest reverse-flow speed and vortex formation length, and is unaffected by the geometry of Rankine oval.

\section{Conclusion}
In this study, we conducted a systematic investigation of the fluid dynamics around Rankine ovals using extensive direct numerical simulations, elucidating how flow regimes, vortex behavior, and force coefficients evolve with the Reynolds number ($Re$) and the parameter ($Ua/m$).
This work enhances our understanding of vortex dynamics, flow separation, and wake characteristics in Rankine oval and bluff body flows.

A primary discovery is the approximately linear relationship between the critical Reynolds number and $Ua/m$, which delineates the onset of vortex shedding.
We also examined the drag coefficient and Strouhal number across a wide range of Reynolds numbers, revealing their dependencies on viscosity and $Ua/m$.
Decomposing the drag into pressure and friction components provided clearer insights into their respective roles in the total drag force.

Furthermore, we analyzed the interplay between vortex formation length, wake width, and Reynolds number.
Through data-driven dimensional analysis, we distilled the governing nondimensional parameters for key flow quantities, uncovering that the friction drag coefficient is primarily determined by $Re_l$ ($Re_l$ using the oval's half-length l), yielding a robust power-law correlation independent of $Ua/m$.
Similarly, the Strouhal number St is dominated by $Re_r$.
These insights simplify the prediction of drag and vortex shedding dynamics across diverse configurations.

Notably, for Rankine ovals with sufficiently large $Ua/m$, the total drag can be reliably approximated without resource-intensive simulations. This is achieved by estimating pressure drag using inviscid potential flow solutions, then combining it with friction drag derived from dimensional analysis for high-fidelity predictions.

Overall, these findings offer valuable insights into the fluid dynamics of Rankine ovals, with broader implications for fundamental fluid mechanics and engineering applications.
Future studies can build upon this framework by applying it to more complex geometries inspired by Rankine ovals, such as those involving multiple source-sink pairs or asymmetric profiles, to further elucidate transitional flow behaviors. 
Additionally, exploring the influences of high Reynolds number flows, unsteady incoming flows, and three-dimensional effects could yield deeper insights into wake instabilities and bluff body flows.

\section{Acknowledgements}
This work is supported by the NSFC Outstanding Research Groups for Multiscale Problems in Nonlinear Mechanics (Nos. 12588201), the National Natural Science Foundation of China (Nos.92252203 and 12102439), the Chinese Academy of Sciences Project for Young Scientists in Basic Research (YSBR-087), and the Strategic Priority Research Program of Chinese Academy of Sciences (XDB0620102).
\section{Declaration of interests}
The authors report no conflict of interest.
\appendix
\section{Derivation of Linear Approximation for Aspect Ratio in Rankine Oval}
Notably, in Table~\ref{tab1}, the relationship between $Ua/m$ and $l/h$ can be approximated linearly.
This approximation is derived for the case where $h=0.5$ as follows:
\begin{equation}
\begin{gathered}
  \hat{x}  = \frac{x}{a},\hat{y}  = \frac{y}{a},C = \frac{{Ua}}{m} \hfill \\
  {{\hat{x} }^2} + {{\hat{y} }^2} - {1^2} = \frac{{2\hat{y} }}{{\tan (2\pi C\hat{y} )}} \hfill \\
  \hat{x}  = 0 \Rightarrow {{\hat{y} }^2} - {1^2} = \frac{{2\hat{y} }}{{\tan (2\pi C\hat{y} )}} \hfill \\ 
\hat{y}  = 0 \Rightarrow {\hat{x} ^2} - {1^2} = \frac{1}{{\pi C}}.\hfill \\ 
\end{gathered} 
\end{equation}
When $\hat{x}  = 0$, we have $\hat{y}  = \frac{0.5}{a}$.
From this, we obtain the following equation:
\begin{equation}
\label{EqC}
C = \frac{a}{\pi } \arctan \frac{a}{{0.25 - {a^2}}}.
\end{equation}
When $\hat{y}  = 0$, we have 
\begin{equation}
\label{Eql}
l = a\sqrt {1 + \frac{1}{\pi C}}.
\end{equation}
The Taylor expansions of Eq.~\eqref{EqC} and ~\eqref{Eql} are given by
\begin{align}
l\sim0.5 + \frac{4}{3}{a^2} + O({a^4})\\
C\sim\frac{4}{\pi }{a^2} + O({a^4}).
\end{align}
Thus, for $h=0.5$, we have the following approximation of aspect ratio:
\begin{equation}
\frac{l}{h} \approx 1 + 2\frac{{Ua}}{m}.
\end{equation}

\end{document}